\pdfoutput=1
\documentclass[12pt]{article}

\usepackage{amsmath}
\usepackage{amssymb}

\usepackage{graphicx}

\usepackage{cite}
\usepackage{chapterbib}

\usepackage{balance}

\usepackage{color} 

\usepackage{setspace} 
\usepackage[labelfont=bf,labelsep=period,justification=raggedright]{caption}

\makeatletter
\renewcommand{\@biblabel}[1]{\quad#1.}
\makeatother

\date{}

\begin{document}

% Title must be 150 characters or less
\begin{flushleft}
{\Large
\textbf{BioFET-SIM Web Interface: Implementation and Two Applications}
}
% Insert Author names, affiliations and corresponding author email.
\\
Martin R. Hediger$^{1\ast}$, 
Jan H. Jensen$^{1}$,
Luca De Vico$^{1}$
\\
\bf{1} Department of Chemistry, University of Copenhagen, Universitetsparken 5, DK-2100 Copenhagen, Denmark
\\
%\bf{3} Author3 Dept/Program/Center, Institution Name, City, State, Country
%\\
$\ast$ Corresponding Author, Email: martin@chem.ku.dk
%\bf{2} Author2 Dept/Program/Center, Institution Name, City, State, Country
\end{flushleft}

%***************************************************************************
% Please keep the abstract between 250 and 300 words
\section*{Abstract}
We present a web interface which allows to conveniently setup calculations based on the BioFET-SIM model.
With the interface, the signal of a BioFET sensor can be calculated depending on its parameters, as well as the signal dependence on pH.
As an illustration, two case studies are presented.
In the first case, a generic peptide with opposite charges on both ends is inverted in orientation on a semiconducting nanowire surface leading to a corresponding change in sign of the computed sensitivity of the device.
In the second case, the binding of an antibody/antigen complex on the nanowire surface is studied in terms of orientation and analyte/nanowire surface distance.
We demonstrate how the BioFET-SIM web interface can aid in the understanding of experimental data and postulate alternative ways of antibody/antigen orientation on the nanowire surface.

%***************************************************************************
% Please keep the Author Summary between 150 and 200 words
% Use first person. PLoS ONE authors please skip this step. 
% Author Summary not valid for PLoS ONE submissions.   
\section*{Author Summary}

%***************************************************************************
\section*{Introduction}
%\textcolor{red}{
A bionanosensor is most generally described as a device that allows the detection of an analyte (e.g. H$^+$ ions, small molecules, proteins, DNA, viruses, cells) at ambient conditions where the dimensionality of the sensitive component is on the nanometer scale.
The sensitive component can be either a functionalized nanotube, nanoribbon or nanowire, the latter being the focus of this paper.
Currently, a large research effort is dedicated to the development and application of bionanosensors including pH measurement\cite{chen2006silicon}, protein sensing\cite{cui2001nanowire, stern2007label, gao2009subthreshold, C0LC00605J}, DNA detection\cite{wong2009directed, nn301495k}, blood analysis\cite{nn2035796}, nanotechnology based medicine\cite{berthing2009applications}, and the description of fundamental performance limits of these sensors\cite{nair2006performance, nair2008screening, nair2010theory}.
A number of reviews describe the bionanosensor\cite{patolsky2006nanowire, waleed2007study, curreli2008real, neizvestny2009semiconductor, roy2009nanostructure} and its components.
In addition to the experimental work, simulators of bionanosensors are being developed and several numerical models have been presented\cite{heitzinger2007computational, heitzinger2008modeling, chen2008charge, heitzinger2009modeling, windbacher2010biotin}.\\
Most simulators are aimed at providing a measure of the current or conduction through the sensor, which are the prime experimental targets.
This requires, in principle, the description of the charge distribution on the sensor and within.
From the charge distribution, the potential within the sensor is calculated which in turn is required for the calculation of the current.
The calculation of the potential can be either numerical or analytical.\\
In this paper, we present a computational tool to simulate a bionanosensor which is based on an analytical model\cite{sorensen2007screening, de2011quantifying, de2011predicting} and which can calculate the sensitivity of the nanosensor and the pH dependence of the signal upon binding of a protein.
The use of an analytical model is mainly motivated by the fact that this model does not require extensive computations but still allows to gain a qualitative understanding of the biosensor problem in a straightforward manner.\\
Furthermore, we have demonstrated\cite{de2011predicting, de2011quantifying} that 1) the experimental data can be reproduced with sufficient accuracy to help interpret them and 2) going beyond the simplifications inherent in the model may not be warranted until the key properties of current BioFET experimental set-ups are known with greater precision.
We note that the presented method, which we refer to as \textit{BioFET-SIM}, has gained popularity in the biosensing community and is being actively incorporated into present day research\cite{C2NR12038K, ja205684a, duan2012quantification, hakim2012thin}.\\
Because of the reduced required computational effort, it is possible to incorporate the model into a browser based application which by doing so can be made accessible to a wide range of users.
Our goal is to provide a tool from which indications for trends in predictions can be obtained with minimum effort of preparation and time.
To further improve the usability, the model is coupled to an atomic representation of the protein structure in a way many researchers in the biocomputational field are familiar with.
Such an application is an ideal tool for gaining insight and obtaining semi-quantitative solutions to the problems at hand which can be of valuable guidance in the design process of an experiment, for optimization of experimental parameters and rationalization.\\
We relate our application to other simulators where we point out the \textit{BioSensorLab}\cite{2929}, which implements settling time, sensitivity and selectivity of the biosensor, \textit{Nanowire}\cite{1307}, which allows to carry out self-consistent three dimensional simulations of a silicon nanowire or \textit{Medici}\cite{medici}, a commercial simulator.
Custom prepared simulators\cite{heitzinger2007computational} have also been described.
To the best of our knowledge, out of all available simulators targeted at modeling of biosensors, the tool we present in this paper is the first to combine a three dimensional visual representation of the biomolecule to be studied directly in the browser with a method to solve the biosensor problem.
%} % textcolor end

The primary use of our tool is to model the binding of proteins to nanowire surfaces for which the structure is available in the PDB database.
However, using a custom prepared structure, it is also possible to model binding of an antibody/antigen complexes\cite{ja205684a}, an illustration of which is provided in the results section.
The authors further envision the application of the program to the modeling of DNA binding.

To put the use of this application into context, we note that every program in general requires a certain amount of preparation.
When using atomic detailed structures, the molecular structure of the pH dependent charge distribution has to be generated, which usually involves the combined usage of a number of different software tools, each dedicated to a particular task.
If different orientations of the analyte charge distribution are to be studied, the procedure needs to be repeated for each orientation.
Furthermore, for each orientation, the nanowire is covered differently, the evaluation of which requires additional manual effort.

The motivation for the development of the web interface is to eliminate this effort as far as possible.
The effort of assigning partial charges to the amino acids is essentially removed.
Instead, using the web interface, any number of orientations of analyte towards the nanowire surface can be generated within minutes.
Furthermore, the coverage of the nanowire is calculated instantaneously for any orientation of the analyte towards the nanowire surface.
Allowing the user to interactively adjust the orientation of the analyte through a Jmol\cite{hanson2010jmol} applet provides a maximum of visual feedback and allows to generate the coordinates of the charge distribution as straightforwardly as possible.

In addition to the web interface, a command line version of the program is available.
The command line version is used together with a special BioFET-SIM input file written by the interface which can be used to reproduce a given calculation locally.

The BioFET-SIM Online web interface is hosted at www.biofetsim.org, the source code for both the web interface and the command line version of the application is hosted at Github, the URL to the repository is found on the interface page.

%***************************************************************************
% You may title this section "Methods" or "Models". 
% "Models" is not a valid title for PLoS ONE authors. However, PLoS ONE
% authors may use "Analysis" 
\section*{Methods} 

A BioFET nanosensor consists mainly of a semiconducting nanomaterial covered by an oxide layer and a (bio-)functionalization layer.
The device is usually immersed in an electrolyte containing the analyte.
In the following, we describe the implementation of each of these domains in the BioFET-SIM program.

%\textcolor{red}{
\textbf{Theoretical Background}\\
%}\\ % textcolor end
% NANOWIRE
The sensitivity $\Delta G/G_0$, where $\Delta G$ is the difference between the conductance upon binding $G$ and the base conductance $G_0$, of the nanowire is evaluated using a Thomas-Fermi screening model for the charge carriers in the nanowire\cite{nitzan2002electrostatic, liang2004electrostatic, zhang2009screening}.
%\textcolor{red}{
We point out that in this model, only one type of carrier in the entire nanowire is considered and the nanowire material is assumed to resemble a low density metal.\\
In this context, the description of the electrostatic problem of the biosensor is governed by two major assumptions regarding the carrier concentration which are 1) the carrier concentration is assumed to follow an uniform distribution throughout the wire and 2) it is assumed not to be influenced by the electrostatic potential due to surface charges.
%}\\ % textcolor end
The sensitivity (assuming a $p$-type doped nanowire) is evaluated by
\begin{equation}
\label{eq:sensitivity}
\frac{\Delta G}{G_0} = - \frac{2}{Rep_0}\Gamma \left[\sum_i^m \left( \Gamma_{l_{i,tot}}\sigma_{b_i} \right) \right]
\end{equation} 
where $R$ is the radius of the nanowire, $e$ is the elementary charge, $p_0$ is the hole carrier density in the nanowire and $\sigma_{b_i}$ is the corresponding surface charge density of the charge $i$ on the biomolecule $b$ containing $m$ ionized sites (residues and termini)\cite{de2011quantifying}.
$l_{i, tot}$ is the distance of the discrete charge $q_i$ above the nanowire surface, which will further be discussed below.
$\Gamma_l$ is given by
\begin{equation}
\label{eq:gamma_l}
\Gamma_l\left(\lambda_D\right) = 2\frac{R}{R+l}\left[1+\sqrt{\frac{R}{R+l}}e^{l/\lambda_D}\right]^{-1}.
\end{equation} 
In Eq. \ref{eq:gamma_l}, $l$ is again the distance between the discrete charge $q_i$ and the nanowire surface, i.e $l_{i, tot}$ in Eq. \ref{eq:sensitivity} and $\lambda_D$ is the Debye screening length of the electrolyte/buffer solution (the expression for which is given below).
The expression for $\Gamma$ is found in section \ref{subsec:gamma} { o}f the supporting material but for the purpose of the discussion can be considered a factor with values ranging from zero to unity.
$\Gamma$ and $\Gamma_{l}$ are dimensionless functions quantifying the actual sensitivity of the nanowire ($\Gamma$) and the effect of $\sigma_{b_i}$ ($\Gamma_l$) and arise from the solution to the Poisson equation in cylindrical coordinates given the boundary conditions of the problem\cite{de2011predicting}.
$\Gamma$ depends on both $\lambda_D$ (describing the ionic strength of the buffer) and the Thomas-Fermi screening length $\lambda_{TF}$ (describing the electric field screening within the wire), whereas $\Gamma_l$ depends only on $\lambda_D$.\\ 
The screening model for the wire is a simplification in the sense that possible deactivation of dopants at the surface\cite{bjork2009donor} or the increased dopand concentration near the surface compared to the semiconductor bulk\cite{garnett2009dopant} is solely described by the screening length $\lambda_{TF}$.
For a $p$-type ($n$-type) semiconductor, the screening length $\lambda_{TF}$ is related to the charge carrier density $p_0$ ($n_0$) through
\begin{equation}
\label{eq:lambda_tf}
\lambda_{TF} = \sqrt{\frac{\hbar^2 \epsilon_r \pi^{4/3}}{m^\ast e^2 p_0^{1/3}}}
\end{equation} 
where $\epsilon_r$ is the relative permittivity of the nanowire material and $m^\ast$ is the effective mass of the charge carrier ($p_0$ would be replaced by $n_0$ for an $n$-type nanowire).
%\textcolor{red}{
From Eq. \ref{eq:lambda_tf}, we note that $\lambda_{TF}$ can be interpreted as a measure for the charge carrier density in the nanowire under no applied bias and that therefore this parameter can be used to simulate the effect of the back gate in an experimental setup.
%} % textcolor end
We note that the accuracy of the predicted signal is strongly dependent on the quality of the estimation of the charge carrier density in the wire, thus for best predictivity, this parameter has to be as close to the actual value of the experimental setup as possible\cite{gao2009subthreshold}.\\
%\textcolor{red}{
We further note that the described linearized model is not capable of describing non-linear effects such as inversion mode of operation.
However, the model distinguishes between accumulation/depletion mode of operation by allowing to choose between a $n$- or $p$-type material and different values of $\lambda_{TF}$.
%} % textcolor end

% OXIDE- and BIOFUNCTIONALIZATION LAYER
%\textcolor{red}{
The oxide layer is known from earlier studies\cite{de2011quantifying} to have an important effect on the predicted sensitivity and is a key component of a BioFET sensor.
The gate dielectric is understood to be in part responsible for biosensor degradation due to the incorporation of charges when exposed to solvent (through ion diffusion)\cite{dorvel2012silicon}.
However, in our approach surface charges formed on the oxide layer surface and within are not taken into account.
In other words, only the signal generated by a charged system bound at the surface of the sensor is considered.
The change in signal given by, e.g., a change in pH which can affect the surface charge density of the oxide layer, is considered as background signal.\\
The biofunctionalization layer is currently implemented solely as a distance parameter, providing a measure of the spatial extension of the linker molecule.
Charges on the linker molecules are not considered.
By using the same distance between the surface and the sensed protein for all proteins, we imply that all proteins are binding in one orientation to the nanowire surface.
This is being further discussed below.\\
We note that in principle the surface functionality of the nanowire is non-uniform\cite{cui2001nanowire} and requires a combined description of the pH dependent charge on the linker molecules as well as the oxide where a common description of the charge of the oxide layer is through the site-binding model\cite{yates1974site}.
%} % textcolor end

% ELECTROLYTE and ANALYTE
The influence of buffer characteristics on device performance has been described\cite{stern2007importance, nair2007design} and we note that the electric screening of the analyte by the buffer can have a considerable effect on the predicted signal\cite{de2011quantifying}.
As stated above, the screening of the analyte signal by the electrolyte is implemented through the expression $\Gamma_l$ which depends on the Debye length $\lambda_D = \sqrt{\frac{\epsilon_0 \epsilon_3 k_B T}{2 N_A e^2 I}}$, where $k_B$, $T$ and $N_A$ indicate, the Boltzmann constant, temperature and Avogadros constant, respectively.
The expression for the ionic strength is given by $I = 1/2 \sum_i c_i z_i^2$ where $c_i$ indicates the concentration of ion species $i$ and $z_i$ is its formal charge.  
Furthermore, $\epsilon_0$ and $\epsilon_3$ denote the free space dielectric constant and the relative permittivity of the electrolyte, respectively.\\
%\textcolor{red}{
The description of the electrolyte by the given approach assumes 1) that the electrolyte is in equilibrium, i.e. the chemical potential is at a minimum and 2) that the value for $\lambda_D$ used in the expression for $\Gamma_l$, Eq. \ref{eq:gamma_l}, is equal to the Debye length of the electrolyte.
We note that in principle these values can differ due to the biofunctionalization layer\cite{sorensen2007screening}.
%} % textcolor end

% ENZYME PROTONATION STATES
The enzyme protonation states are described classically.
Depending on the pK$_a$ value, the charge on residue $i$ is calculated as a function of pH using Eq.~\ref{eq:charge_pH} 
\begin{equation}
q_i(pH) = \frac{10^{pK_a^i - pH}}{1 + 10^{pK_a^i - pH}} - p(i)
\label{eq:charge_pH}
\end{equation} 
where $p(i) = 1$ for $i \in$ \{Asp, Glu, C-, Tyr, Cys\} and $p(i) = 0$ else (the charge is evaluated only for ionizable residues).
In Eq. \ref{eq:charge_pH}, $q_i(pH)$ can be interpreted as the probability of the amino acid being protonated\cite{ullmann1999electrostatic}.
The three-dimensional protein charge distribution is obtained from placing the charge calculated from Eq. \ref{eq:charge_pH} at the average of the coordinates of the terminal atoms of the side chain of residue~$i$.
%\textcolor{red}{
The charges of the enzyme residues are calculated solely depending on the pH of the electrolyte and their respective pK$_a$ values as computed by PROPKA.
Binding to the nanowire is assumed not to affect these pK$_a$ values nor to disrupt the overall protein conformation.
%} % textcolor end

%\textcolor{red}{
\textbf{Interface Operation}\\
%}\\ %textcolor end
The interface is shown in Fig. \ref{fig:int}. The interface operation is grouped into three steps: 1) Initialization, 2) Jmol based calculation setup and 3) BioFET-SIM-signal/pH-response calculation.

\textbf{Initialization, Fig. \ref{fig:int}A.}
%\textcolor{red}{
On loading the interface, the user is requested to grant access to the client computer by the Java applet.
This is required if the user wants to be able to save a Jmol state file or to restore a previous session.\\
%}\\ % textcolor end
The calculation is prepared by setting the PDB identifier and the pH value.
Alternatively, the user can upload a custom made molecular structure (in PDB format), which is then being submitted to the web interface.
In case the user uploads a custom prepared PDB file to the web interface, this PDB file has to contain the \verb+MODEL+ and \verb+END+ tags, a generic example is provided in section \ref{sec:sup_biof_comd} { o}f the supplementary material.  
%\textcolor{red}{
After successfully uploading a PDB file, the structure can be loaded into the interface by using its file name (without extension) in the PDB identifier input field.\\
%}\\ % textcolor end
The following steps are carried out in the background by clicking ``Initialize''.
The server first checks the availability of the requested PDB file in an internal database (assuming no file was uploaded) and downloads the PDB file of the biological assembly from the PDB database\cite{berman2000protein} (www.pdb.org) if needed.
The file is processed using PDB2PQR v1.7\cite{dolinsky2004pdb2pqr, dolinsky2007pdb2pqr} to fix any missing side chain atoms.
The structure is realigned to its main rotational axes and its center of mass is placed at the coordinate origin using the VMD\cite{humphrey1996vmd} packages \verb+ORIENT+ and \verb+la1.0+.
The pK$_a$ values of the ionizable amino acids are computed using PROPKA v3.0\cite{ct100578z}.
Since ligand molecules are discarded from the PDB file during the preparation of the calculation, the additional computational effort of calculating the pK$_a$ values by PROPKA v3.1\cite{ct200133y} can be avoided.
The C-terminus is added by the PDB2PQR routine (in form of an \verb+OXT+ atom), while the backbone nitrogen of the first amino acid of each chain represents the N-terminus.  
In order to display the generated discrete charge distribution, the charges and the respective coordinates are written to a PQR file where atomic radii are arbitrarily set to 1.0\AA.
This PQR file thus contains only as many entries as there are ionizable residues and backbone termini present in the biomolecule.
After carrying out these steps, the structure is loaded into the Jmol applet.
The CPU time required to carry out all of the above described steps depends mostly on the size of the molecule.
On average, a time of 1-2 minutes is observed for a PDB file representing a medium sized protein (around 300 residues).
The most time demanding step is the realignment of the structure to the coordinate axis.
However, if a PDB identifier is selected for which the aligned structure is already present on the server, the realignment step is skipped and the time requirement is significantly reduced.
%\textcolor{red}{
Using the ``Reinitialize'' checkbox, the interface can be instructed to carry out all previous steps even if a structure with the same name is already present on the server.
This is required if a file is uploaded for which an older version with the same name is already present on the server.
%} % textcolor end

\textbf{Calculation setup, Fig. \ref{fig:int}B, C.} A Jmol representation of the computed charge distribution overlayed with a ribbon representation of the biomolecule is displayed.
A flat plane of carbon atoms illustrates the nanowire surface (without having any influence on the computed results).
In this Jmol applet, the user can adjust the orientation of the biomolecule towards the nanowire surface allowing to take into account how the biomolecule binds according to the position of its binding sites.
Also, it is possible to study the effect of different orientations on the signal, in particular if a specific orientation has a significantly different signal compared to other orientations.  
The parameters (Fig. \ref{fig:int}C) defining the BioFET-SIM calculation can be adjusted below the Jmol applet and they correspond to the parameters introduced in Table \ref{tab:parameters}.
%\textcolor{red}{
Recommended lower and upper limits for the parameters, as well as a tool to calculate the charge carrier density from $\lambda_{TF}$ (Eq. \ref{eq:lambda_tf}), is provided on a separate help page, the link for which is found on the interface.\\
%}\\ % textcolor end
The BioFET-SIM calculation requires the computation of the normal distance ($z$-coordinate) of the discrete charges from the nanowire.
Since the structure is placed at the coordinate origin, the atoms and charges have formally positive and negative $z$-values in the Jmol applet (Fig. \ref{fig:distance_refs}A).
When submitting the calculation by clicking the "BioFET-SIM" button, internally all charges are offset by the most negative $z$-value, $z_{min}$, Fig. \ref{fig:distance_refs}B.
Due to the offsetting of the coordinates, any free space between the biomolecule and the nanowire introduced by adjusting the orientation has no effect on the computed results.
Together with the biolinker- and oxide layer thickness, the total distance, $l_{i, tot}$, of each discrete charge from the nanowire surface is computed (Fig. \ref{fig:distance_refs}C) and used in the evaluation of the sensitivity by Eq. \ref{eq:sensitivity}.\\
Upon submission of the calculation to the server, the number of biomolecules covering the nanowire in the given orientation is determined by dividing the nanowire surface with the area of the face of the bounding box of the biomolecule oriented towards the nanowire, Fig. \ref{fig:coverage}.
In doing so, it is assumed that the nanowire is completely covered by biomolecules, that all biomolecules are oriented in the same way and, as stated above, that all bound biomolecules are equally distant from the nanowire surface.
%\textcolor{red}{
This is reasonable to assume, when considering high affinity binding between as e.g. in the biotin and (strept-)avidin complexes\cite{diamandis1991biotin}.
%} % textcolor end
Complete coverage of the nanowire has been demonstrated experimentally\cite{liu2010specific}.
Alternatively, the web interface also allows the manual setting of a parameter defining the number of proteins covering the nanowire surface independent of the orientation of the biomolecule or the nanowire surface area.
This feature is added to the web interface because it is questionable if the number of molecules should adjust with orientation or not.
For non-globular proteins, the required area on the surface can vary strongly with orientation, however the number of linker molecules is assumed to remain the same for two different orientations.\\
By selecting the ``Single'' option, the web interface also allows to use the single charge model\cite{de2011quantifying}, where the overall charge of the analyte is placed at the geometrical center of the enclosing bounding box and the discrete charge distribution within the protein is not considered explicitely.
%\textcolor{red}{
The single charge mode of interface operation is useful when no particular binding orientation is favoured.\\
When a calculation has been carried out, a Jmol state file can be saved on the user machine.
This file allows to restore a session at a later point in time.
As stated above, this option is only available if the user grants access to the signed applet, else the state file can not be written to the user machine.
We demonstrate the restoration of a session in an instruction video (URL is found on the interface page).
%} % textcolor end

\textbf{Calculation of results, Fig. \ref{fig:int}D.}
%\textcolor{red}{
Two types of calculations can be performed:
\begin{enumerate}
\item BioFET-SIM signal, giving the sensitivity as a function of the parameter selected using the "Plot" radio button in the indicated range (this calculation type is illustrated in the discussion of the generic peptide model)
\item pH response, giving the sensitivity as a function of pH for the parameters entered (shown in Fig. \ref{fig:int}D)
\end{enumerate}
%} % textcolor end
The pH response signal is computed by evaluating the BioFET-SIM signal at different pH values which will correspond to different partial charges on the residues of the protein.
The plotted data and a specially formatted input file for the command line version of the BioFET-SIM program can be downloaded after the calculation is carried out.
The input file contains all parameters together with the charge distribution and allows to carry out the calculation with the command line version of the BioFET-SIM program (system requirements and usage instructions are given in supplementary material).
%\textcolor{red}{
For convenience, a label indicates the sensitivity and the base conductance computed at the given set of parameters, Fig. \ref{fig:int}C, top.
%}% textcolor end

\subsection*{BioFET-SIM Command Line Version Description}
The command line version of the BioFET-SIM program can be used to run calculations locally after the orientation of a biomolecule towards the nanowire has been established using the web interface.
In order to do so, a BioFET-SIM input file (with .bfs file extension) containing the charge distribution and the BioFET-SIM parameters can be downloaded from the web interface after running a calculation.
The input file is in binary format and not directly human readable.
However, using the command line version of the BioFET-SIM program, the parameters can be viewed and adjusted. 
The command line version of the program can be used for automated calculations.
The usage of the command line version is illustrated in section \ref{sec:sup_biof_comd} { o}f the supporting material.
The command line version is open source and is hosted at Github (the URL is provided on the web interface page).

%*****************************************************************
% Results and Discussion can be combined.
\section*{Results and Discussion}\label{sec:resl} 
To illustrate the use of the web interface, we perform two case studies. In the first case, a generic linear peptide is placed on the nanowire and the dependence of the sign of sensitivity on the orientation of this peptide is evaluated (Figs. \ref{fig:kk8add}A-C). In the second case, we demonstrate the effect of different orientations of an antibody/antigen complex on a relative sensitivitiy value and relate to experimental work by the Reed group\cite{ja205684a}.

\subsection*{Generic Peptide Model}
The generic peptide used in this study is prepared using the molecular building feature of the PyMOL\cite{PyMOLu} program. The peptide consists of two (protonated) Lys at the N-terminus and two (deprotonated) Asp residues at the C-terminus which are bridged by 8 Ala residues (the termini contribute the third charge at each end of the molecule). The overall charge is -0.23 formal charges at pH 7.4, the nanowire configuration corresponds to the default values as shown in Table \ref{tab:parameters}, the calculation is carried out for a $p$-type nanowire.  

In the orientation of Fig. \ref{fig:kk8add}A the negatively charged aspartic acids are close to the nanowire surface, in Fig. \ref{fig:kk8add}B the positive and negative charges are roughly equally distant from the surface, and in Fig. \ref{fig:kk8add}C the positively charged lysine residues are close to the nanowire surface, respectively.
For each orientation, the dependence of sensitivity on Debye length $\lambda_D$ is computed and shown in Fig. \ref{fig:orient_res}A.

It is clearly visible how the orientation affects the sign of the sensitivity.
When the negative charges on the Asp residues are closer to the wire (Fig. \ref{fig:kk8add}A), positive charge carriers are accumulating in the wire leading to increased conductivity.
When both Asp and Lys residues are equally distant from the wire (Fig. \ref{fig:kk8add}B), the effect on the charge carriers cancels.
When the Lys residues are closest to the wire (Fig. \ref{fig:kk8add}C) the situation is reversed such that positive charge carriers in the nanowire are repelled by the positive charges on the peptide, rendering the nanowire in depletion.
The slightly different absolute values of the sensitivity at a given value of $\lambda_D$ for the two vertical orientations are due to the not exactly inverted orientation of the peptide on the nanowire and due to the fact that the charges are not distributed in a perfectly symmetrical way on the peptide (the Lys side chains being longer than the Asp side chains). This results in slightly different population numbers on the nanowire for the two orientations.

%\textcolor{red}{
The observed signal is further rationalized in terms of the functional form of $\Gamma_l$, Eq. \ref{eq:gamma_l}.
In Fig. \ref{fig:orient_res}B, $\Gamma_l$ is plotted as a function of the charge-surface distance $l_{i, tot}$ for different values of $\lambda_D$.
The plots illustrate that $\Gamma_l$ is comprised in the [1, 0] interval.
When considering the orientation of the generic peptide ($\approx$ 3.8 nm long) reported in Fig. \ref{fig:kk8add}A, the aspartate charges are close to the surface, which means $\Gamma_{l, Asp}$ is close to 1 and contributes significantly to $\Gamma$ through the product $\Gamma_{l_{i,tot}} \cdot \sigma_{b_i}$ (Eq. \ref{eq:gamma_l}).
The lysine side chain charges, instead, are at a distance from the surface for which $\Gamma_{l, Lys}$ is observed to be close to zero.
Therefore $\Gamma_{l_{i, tot}} \cdot \sigma_{b_i}$ of the lysines is minimal.
Only by diluting the buffer solvent (e.g. $\lambda_D = 4.0$ nm) these charges could contribute more to the signal.
%} % textcolor end

\subsection*{Antibody Study}
In the second case study, the web interface is used to study the effect of binding different orientations of an antibody/antigen model complex.
Experimentally, it was shown that different orientations of the antibody are responsible for different signals, which are indicative of different distances between the charged antigen and the surface of the nanowire\cite{ja205684a}.
Two possible binding states of the antibody appear plausible. In one state, the antibody is bound by an N-terminus which is located on the antigen-binding fragment (Fab), Fig. \ref{fig:ab_complex}.
In the other state, the antibody is bound to the nanowire surface by one or both C-termini at the base of the antibody.
When binding through the N-terminus, the antigen is reported to bind at a distance of 5.9$\pm$0.6 nm and when binding through the C-termini, the antigen is reported to bind at a distance of 8.4$\pm$0.4 nm above the nanowire surface\cite{ja205684a}.

Using the web interface, different orientations of the antibody/antigen complex have been generated for both states and studied in terms of their effect on sensitivity.
A description of the preparation of the molecular model of the antibody/antigen complex used for the study and molecular images of the different orientations are provided in section \ref{sec:sup_ab_prep} and Fig. \ref{fig:orientations} { a}nd the raw data is reported in Tables \ref{tablerawdata} { a}nd \ref{tabletofit} {  o}f the supplementary material.
%\textcolor{red}{
We note that for the purpose of this study, the antibody is considered not to interact electrostatically with the nanowire.
It is introduced merely as an advanced form of spacer and as a guidance in the construction of different orientation schemes for the binding.
%} % textcolor end

The two states are characterized by different degrees of freedom to orient the antibody/antigen complex on the surface. For both states, the considered orientations are defined in the schemes of Fig. \ref{fig:ab_orientations}.

The orientations A-C correspond to the orientations considered in the experimental work by Reed et al\cite{ja205684a}. The orientations D-G were generated while considering further possible binding orientations under the conditions imposed by the binding through N- or C-termini.

In the orientations A, C and G, the C-termini at the base of the antibody restrict the movement of the antibody with respect to the nanowire. In contrast, when binding through the N-termini on the Fab, the antibody is more free in its movement on the nanowire surface and the antigen can be placed at a larger range of different distances. These orientations are indicated by the schemes B, D, E and F.

Following the derivation provided in section \ref{sec:sup_derv_sens} { o}f the supporting material, the average distance between the antigen and the nanowire surface, $l$, can be estimated by fitting the expression for the relative sensitivity factor, $\Gamma_l/\Gamma_l^{max}$, for different values of $\lambda_D$.
For the orientations A-G of the antibody bound in one of the two states, the computed values of the relative sensitivity factor are shown in Figs. \ref{fig:ab_data}A and B together with the value of $l$ obtained from the curve fit.

Considering the limited movement of the antibody when binding by the C-termini, the antibody is required to remain upright on the nanowire surface. For the orientations A, C and G, $l$ is found to be between 9.8 and 14.5 nm. For the state bound by the N-terminus on the Fab (giving rise to the orientations B, D, E and F), $l$ is found to be in the range of 5.9 to 17.8 nm. In this range, the lowest two values of $l$, 5.9 and 7.9 nm, correspond to orientations in which the antibody is lying on the surface of the nanowire (orientations E and F). The molecular image of the antibody orientation corresponding to curve l$_F$ is shown in Fig. \ref{fig:ab_if}.

The orientation F results in a computed relative sensitivity factor for which the fitted value of $l$ is in best agreement with the antigen/nanowire surface distance of around 5.9 nm reported in the experiment. 

From this case study, it is postulated that in addition to the orientations considered so far by Reed et al., a number of other orientations appear plausible as well.
Based on our findings, we postulate that orientations E or F of the antibody are most likely to explain a signal corresponding to an antigen/nanowire distance of 5-7 nm and an upright position (A, G) is most likely to explain a signal corresponding to an antigen/nanowire distance of 10-13 nm.
In addition, we observe that the orientation C is likely to correspond to an orientation where the antigen is placed even further away from the nanowire surface and is thus unlikely to explain the experimentally observed low value of $l$.

\subsection*{Conclusions}

We describe a web interface to model the signal of protein binding to a nanowire based BioFET sensor.\\
%\textcolor{red}{
In the model, the nanowire is described using Thomas-Fermi theory, assuming uniform carrier distribution of one carrier type and no deactivation of dopants.
The oxide layer is described through its thickness and permittivity, without considering surface or buried charges.
The biofunctionalization layer is considered to provide a distance measure of the analyte to the nanowire surface, however it is not considered as carrying charges and is assumed to bind all analytes identically.
The electrolyte is described using Debye theory assuming equilibrium conditions.
The charge distribution on the analyte (protein) is calculated from PROPKA and is assumed not to be influenced by the binding to the nanowire surface.\\
We point out that the presented method is considered a tool which can provide qualitative insight into the biosensor problem, especially in cases where not all key experimental parameters are available\cite{de2011quantifying}.
%} % textcolor end

The web interface presented in this work enables efficient and convenient use of the BioFET-SIM model.
The automated generation of the pH dependent charge distribution and the freely rotatable 3D representation of the biomolecule allow to study the effect of geometrical orientation and charge distribution on the sensitivity.
By providing these features, the web interface significantly reduces the previously required manual effort of preparing a BioFET-SIM calculation.
In addition, the web interface is platform independent making it possible to use the BioFET-SIM model within any operating system environment and requiring only a Java enabled web browser being installed on the local machine.

A specially formatted input file prepared by the web interface allows to redo a calculation using the command line version of the BioFET-SIM program locally.

For studying less complex systems consisting of only one formal charge (e.g. binding of glutamate), it is also possible to use the previous version of the web interface.

Two applications of the web interface are illustrated.
In the first, the change in sign of the sensitivity is demonstrated using a generic linear peptide model with opposing charges on each end.
In the second application, the web interface is used to study the binding of an antibody/antigen complex.
A number of orientations are studied and we use the web interface to interpret experimental data published by Reed et al\cite{ja205684a}.
Based on the findings, it is concluded that the previously postulated orientation of the antibody/antigen complex is not necessarily the most reasonable explanation of the observed signal.
It is postulated that an orientation where the antibody/antigen complex is lying on the nanowire surface, is most appropriate to explain the observed value of the antigen/nanowire distance reported by the Reed group.
Furthermore, based on our findings, we rule out one of the proposed orientations as not plausible.

%*****************************************************************
% Do NOT remove this, even if you are not including acknowledgments
\section*{Acknowledgments}
MRH acknowledges the financial support from the Universitetsforskningens Investeringskapital (UNIK) Synthetic Biology program. The authors acknowledge No\'{e}mie Lloret for fruitful feedback on the interface design.

%*****************************************************************
%\section*{References} 
% The bibtex filename
%\bibliography{000-arxiv_bio_main}
\bibliographystyle{001-arxiv_2009}
\bibliography{citations}

%*****************************************************************

\begin{onecolumn}

\section*{Figure Legends}

\begin{figure}[!ht]
\centering
\includegraphics[width=0.80\linewidth]{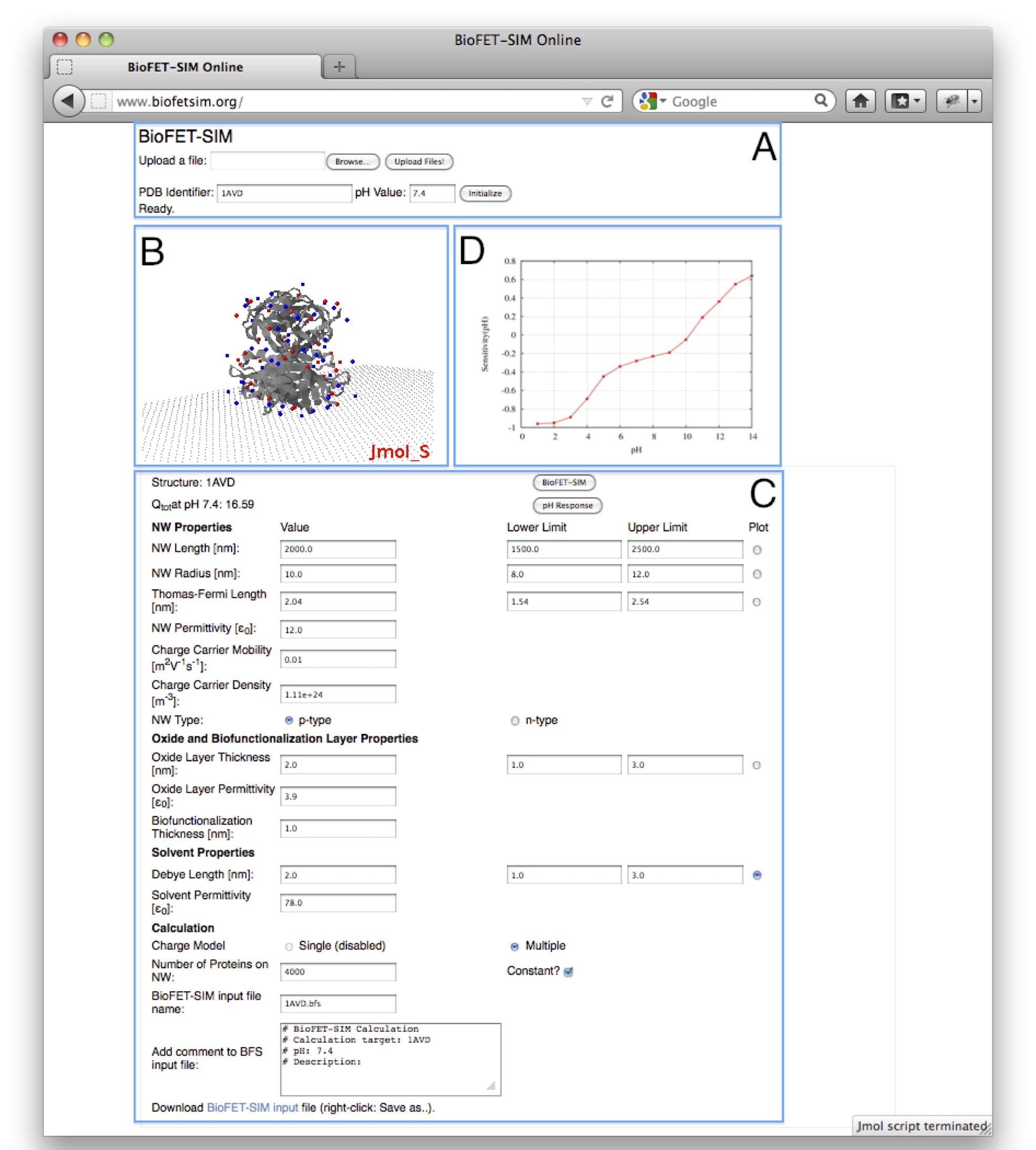}
\caption{{\bf BioFET-SIM Web Interface.} { A:} Upload or request of protein structure and pH setting. B: Jmol visualization of protein on nanowire surface. C: BioFET-SIM parameter section. D: BioFET-SIM calculation result.}
\label{fig:int}
\end{figure}

\begin{figure}[!ht]
\includegraphics[width=1.0\linewidth]{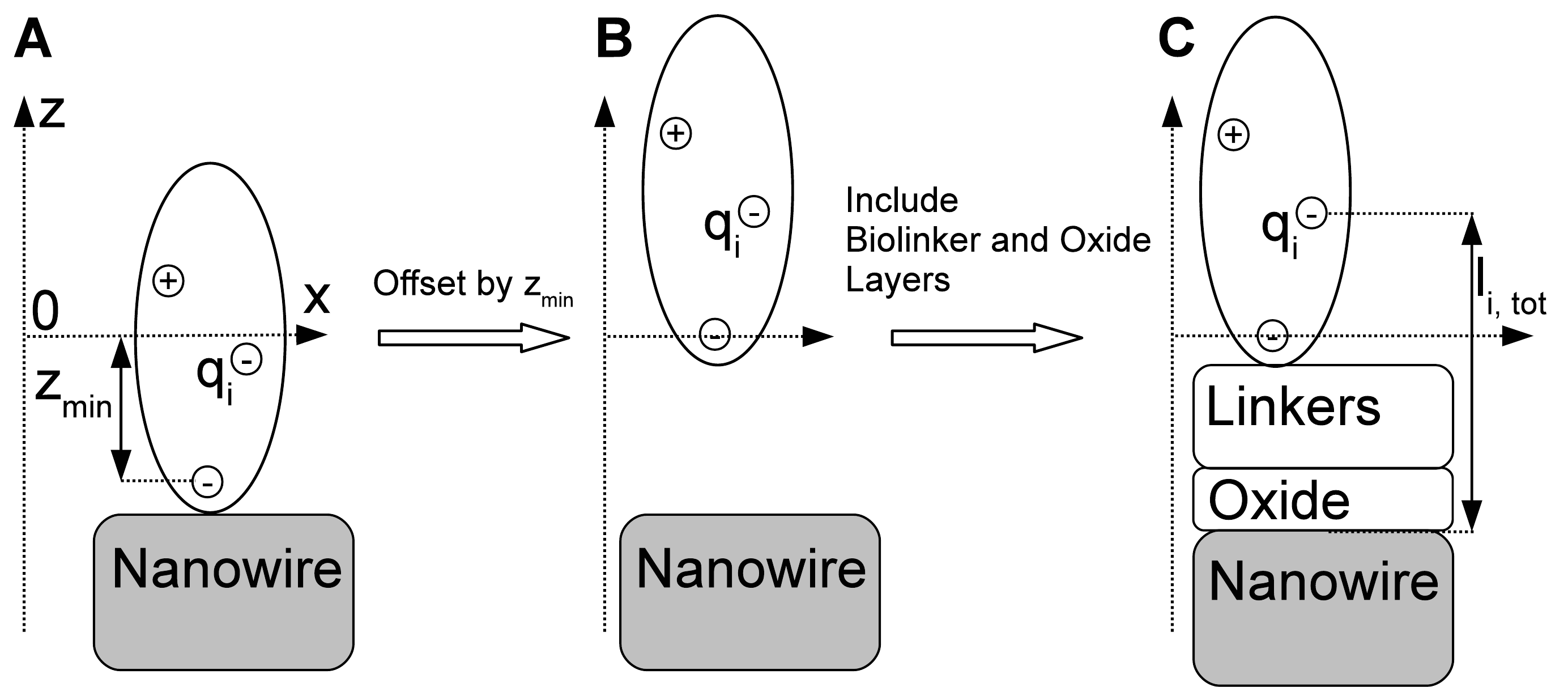}
\caption{{\bf Definition of distance reference system.} A: Protein center of mass
aligned to coordinate origin (z-axis is offset to left for clarity). B: Protein
structure offset by z-min. C: Definition of distance of discrete charge,
$l_{i,tot}$, to NW surface.}
\label{fig:distance_refs}
\end{figure}

\begin{figure}[!ht]
\includegraphics[width=1.0\linewidth]{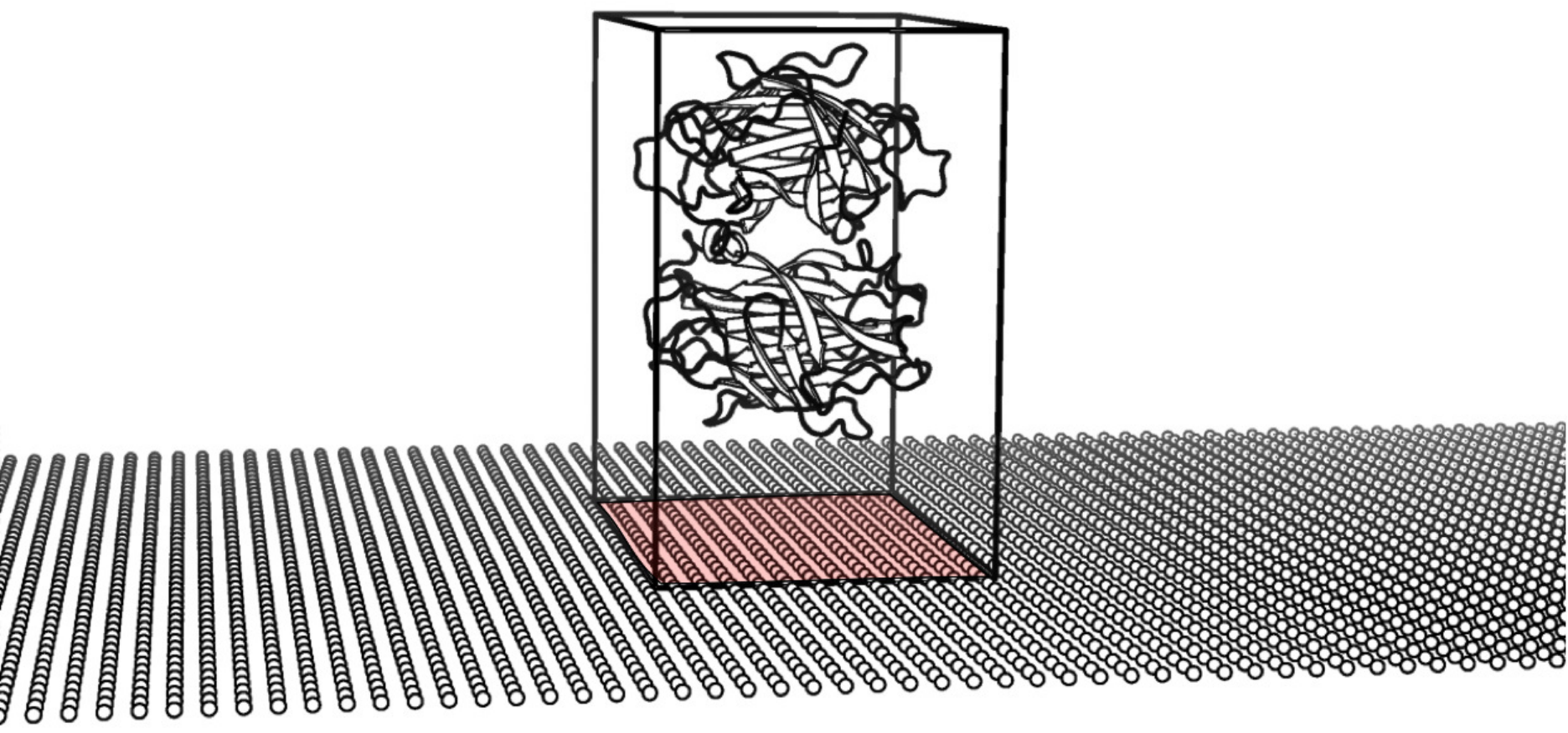}
\caption{{\bf Illustration of occupied surface area on NW.} { R}ed area indicating coverage of the NW by a single biomolecule.}
\label{fig:coverage}
\end{figure} 

\begin{figure*}[!ht]
\includegraphics[width=0.99\linewidth]{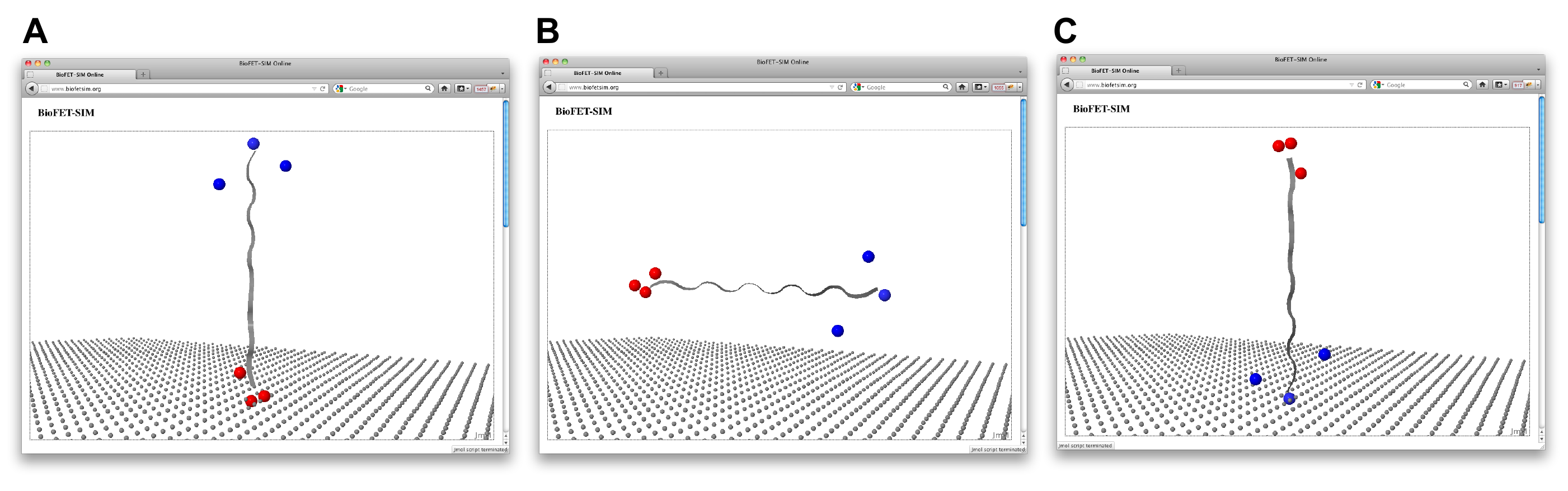}
\caption{{\bf Manual rotation of a generic KK\{8A\}DD peptide in the Jmol applet.} { A}: Asp close to NW, $N$=136976, B: Asp and Lys equally distant from NW, $N$=25462, C: Lys close to NW, $N$=139821.}
\label{fig:kk8add}
\end{figure*}

%\begin{figure}[!ht]
%\centering
%%\includegraphics[width=0.8\linewidth]{018-fig_orie_res}
%\caption{{\bf Dependence of sensitivity on orientation and Debye length.} { T}he color of the data series corresponds to either the Asp or Lys residues being close to the nanowire surface in Figs. \ref{fig:kk8add}A and C. The black data series corresponds to both Asp and Lys residues being equally distant from the surface, Fig. \ref{fig:kk8add}B.}
%\label{fig:orient_res}
%\end{figure}

\begin{figure*}[!ht]
\centering
\includegraphics[width=1.0\linewidth]{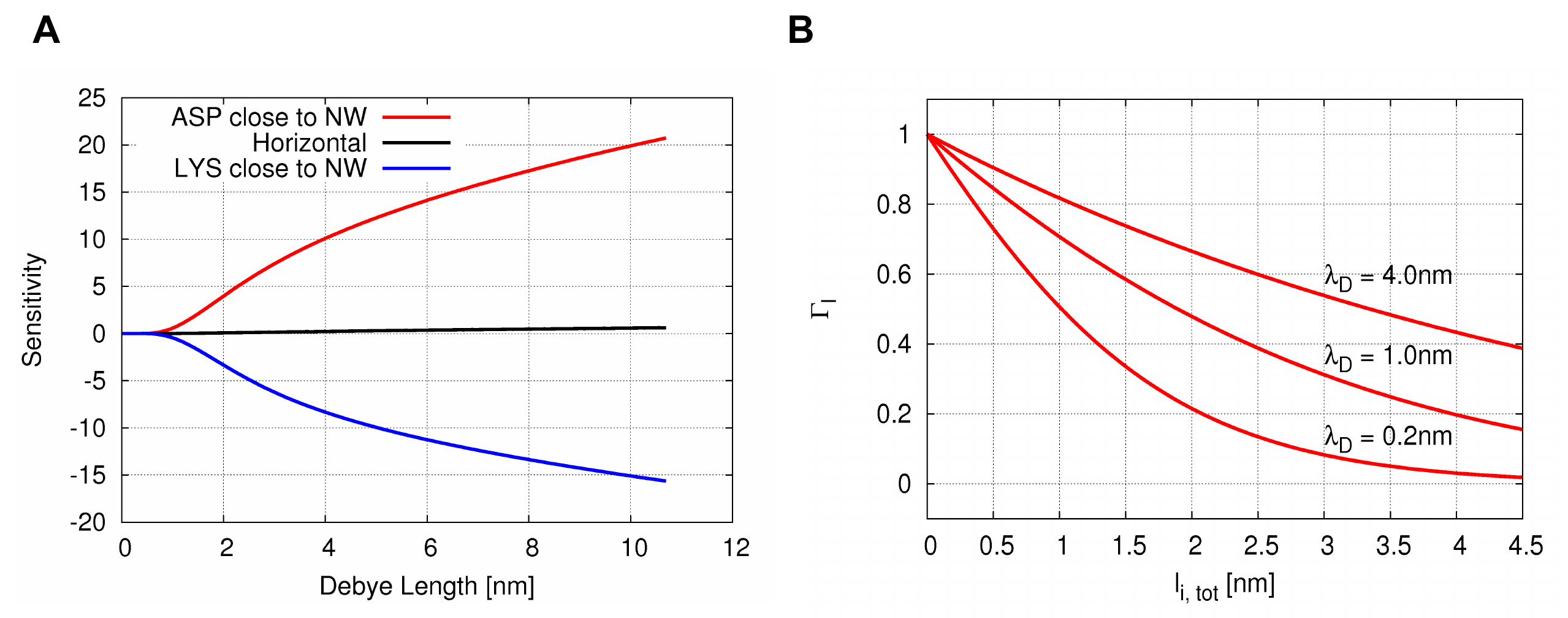}
\caption{{\bf Dependence of sensitivity on orientation and Debye length and $\Gamma_l$ dependence on $l_{i, tot}$.} A: { T}he data series corresponds to either the Asp or Lys residues being close to the nanowire surface in Figs. \ref{fig:kk8add}A and C. The black data series corresponds to both Asp and Lys residues being equally distant from the surface, as in Fig. \ref{fig:kk8add}B. B: Dependence of $\Gamma_l$ on $l_{i, tot}$, Eq. \ref{eq:gamma_l}, for different values of $\lambda_D$ and $R = 10.0$ nm. For $\lambda_D = 0.2$ nm, the function vanishes for $l_{i, tot} > 3.5$ nm.}
\label{fig:orient_res}
\end{figure*}

\begin{figure}[!ht]
\includegraphics[width=1.0\linewidth]{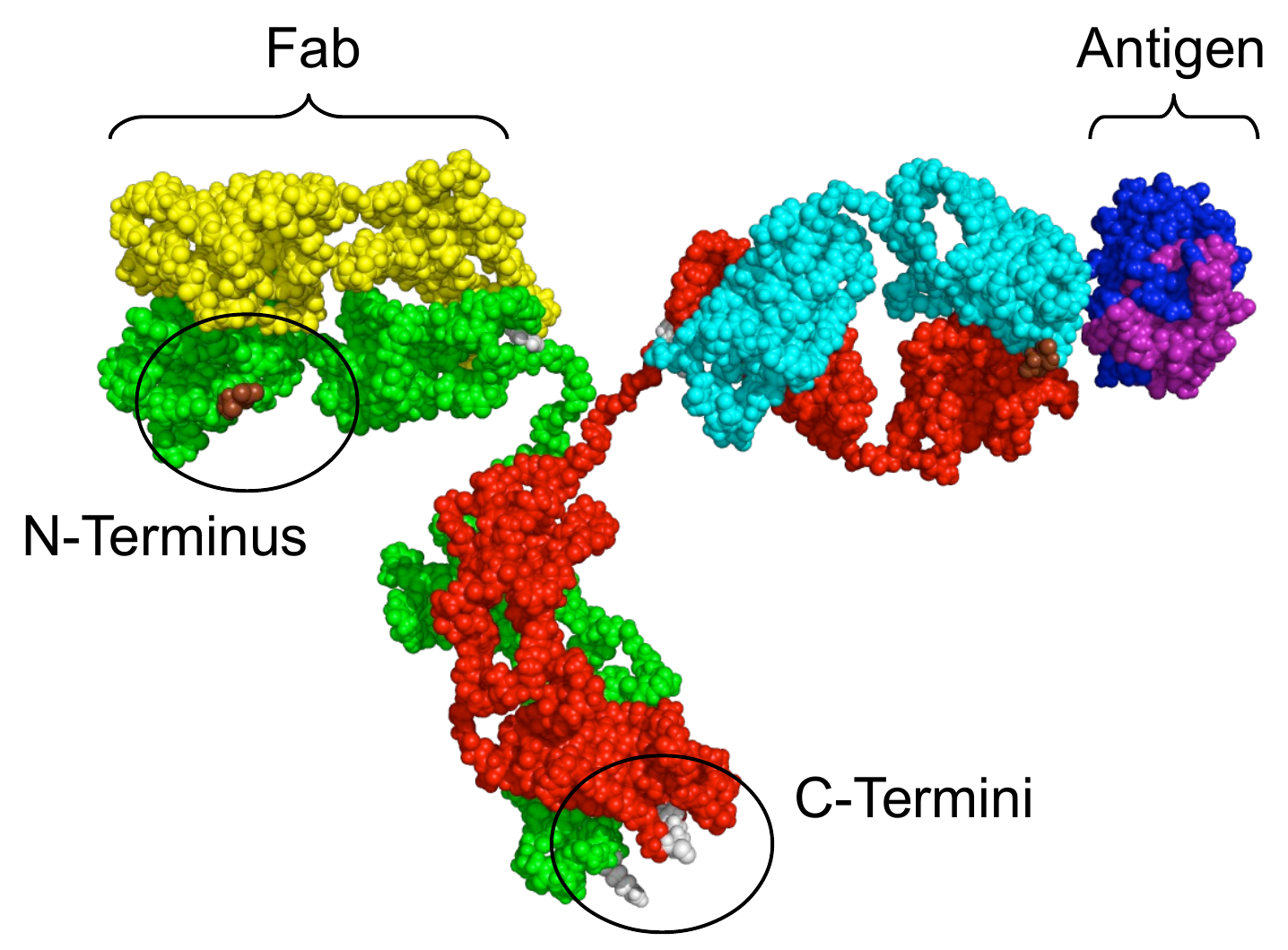}
\caption{{\bf Illustration of custom prepared antibody/antigen system.} { C}-termini in gray at antibody base, N-terminus on Fab in brown.}
\label{fig:ab_complex}
\end{figure}
\newpage

\begin{figure}[!ht]
\includegraphics[width=1.0\linewidth]{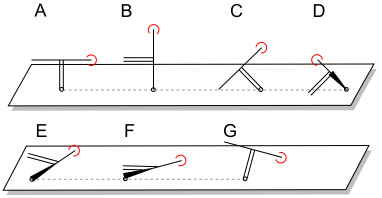}
\caption{{\bf Definition of studied antibody orientations.} { T}he antigen is indicated by a red arc. The antibody base is indicated by a double line. The point of attachement to the NW surface is indicated by the small circle. Orientations A, C, G are bound by the C-termini. Orientations B, D, E, F are bound by the N-terminus.}
\label{fig:ab_orientations}
\end{figure}

\begin{figure*}[!ht]
\includegraphics[width=0.99\linewidth]{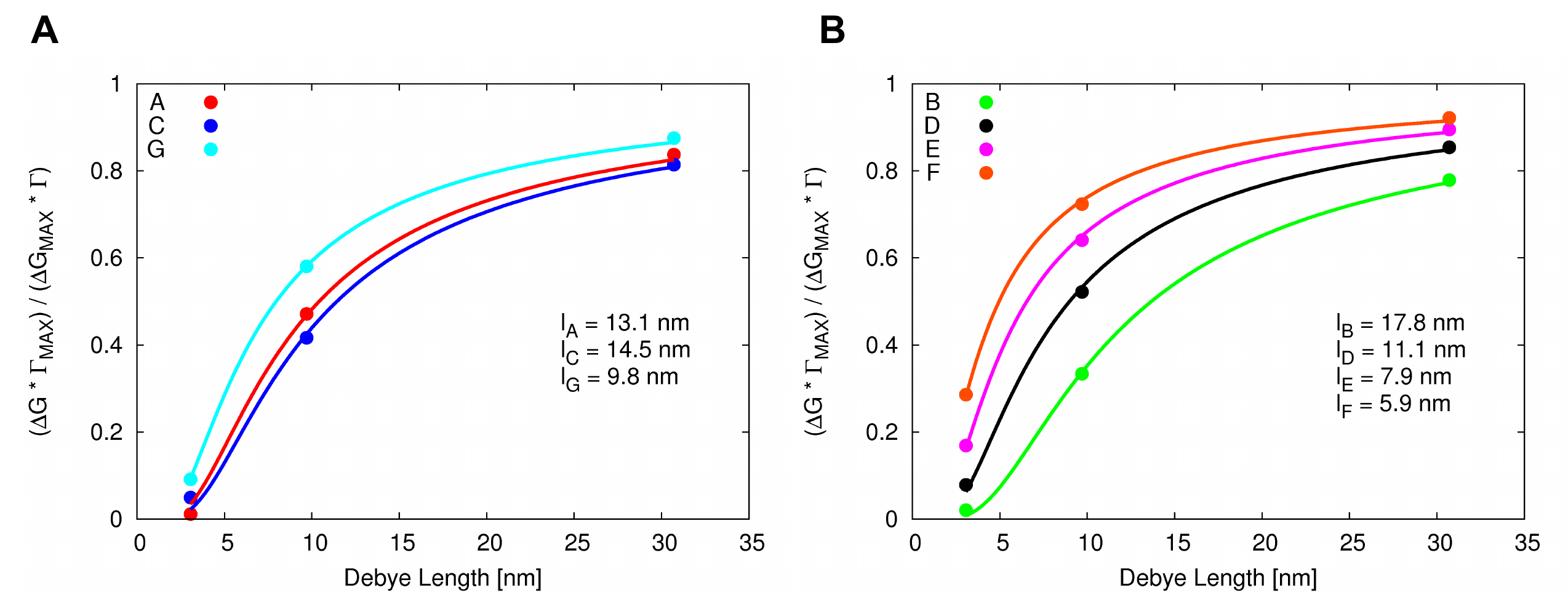}
\caption{{\bf Computed dependence of relative sensitivity factor on Debye screening length.} { A}: Binding by C-termini. B: Binding by N-terminus.}
\label{fig:ab_data}
\end{figure*}

\begin{figure}[!ht]
\includegraphics[width=1.0\linewidth]{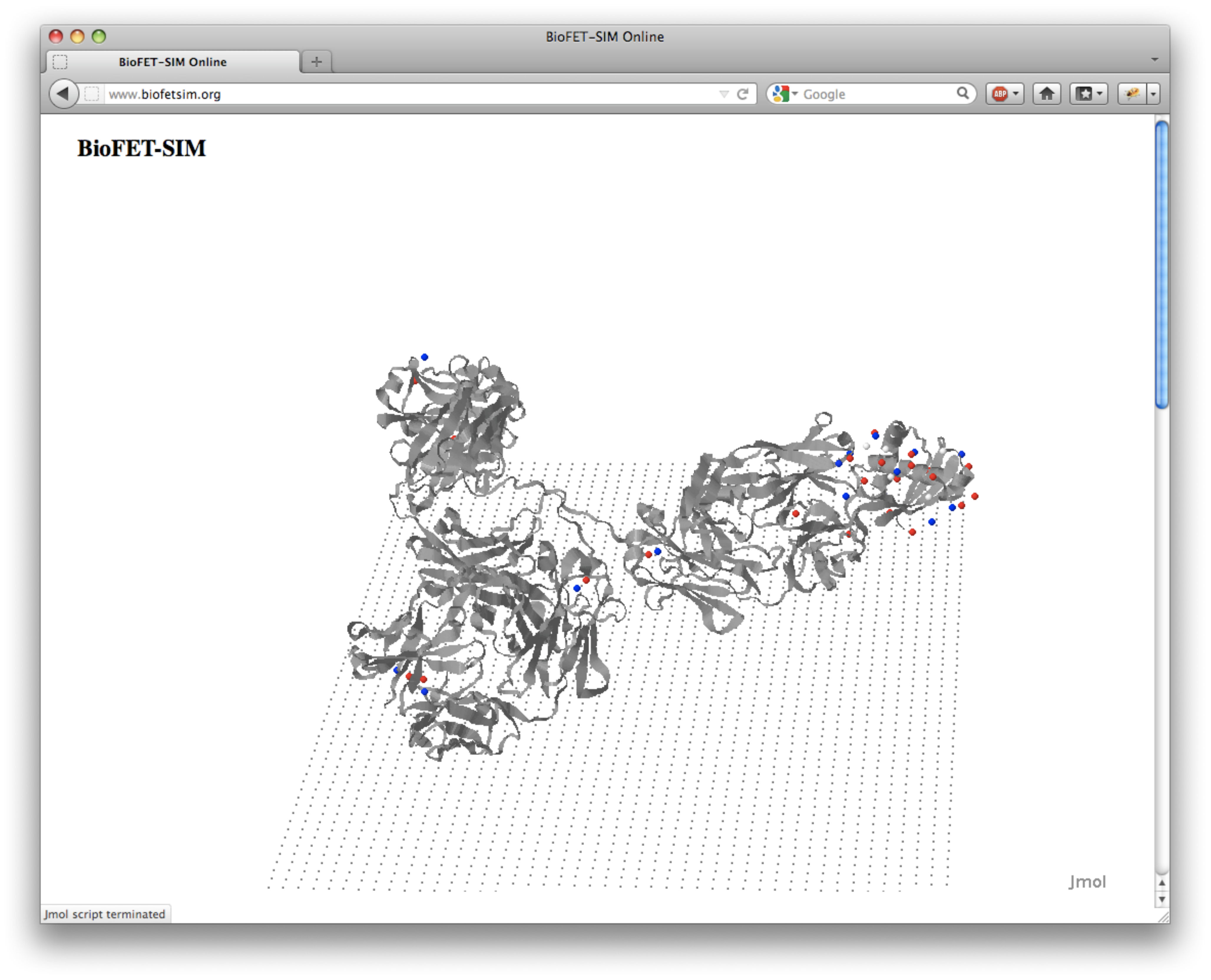}
\caption{{\bf Antibody in orientation F lying on NW surface.} { A}ntigen charge distribution is bound to the right Fab.}
\label{fig:ab_if}
\end{figure}

\clearpage

%%*****************************************************************
\section*{Tables} 
\begin{table}[!ht]
\footnotesize
\begin{tabular}{llrrl@{\extracolsep\fill}}
Domain          & Parameter      & Default & Unit                  & Description\\\hline
NW Properties   & $L_{NW}$       & 2000    & nm                    & NW length\\
                & $R_{NW}$       & 10      & nm                    & NW radius \\
                & $\lambda_{TF}$ & 2.04    & nm                    & Thomas-Fermi screening length\\
                & $\epsilon_1$   & 12.0    & $\epsilon_0$          & NW permittivity\\
                & $\mu$          & 1E-2    & m$^2$V$^{-1}$s$^{-1}$ & Charge carrier mobility,\\
  &  &  &  & $\mu = \frac{\hbar^2}{2m^\ast}(3\pi^2n)^{2/3}$, $n$: electron concentration\\
  & $\kappa_0$ & 1.11E24 & m$^{-3}$ & Charge carrier density, $\kappa \in$ \{$n$, $p$\}, $\kappa:=\kappa(\lambda_{TF})$\\
              & $K$            & $p$     &                     & NW doping type, $K \in \{n$, $p\}$\\\\
Oxide layer and    & $l_{ox}$         & 2.0     & nm           & Oxide layer thickness\\
biolinker properties & $\epsilon_{2}$ & 3.9     & $\epsilon_0$ & Oxide layer permittivity\\ 
                                      & $l_{b}$ & 1.0 & nm     & Biolinker thickness\\\\
Solvent properties & $\lambda_{D}$    & 2.0     & nm           & Solvent Debye length\\
                   & $\epsilon_3$     & 78      & $\epsilon_0$ & Solvent permittivity\\
Biomolecule properties & $N$ & 4000   &                        & Number of biomolecules on NW\\
                &                 &             &   & (computed internally or defined by user)\\\hline
\end{tabular}
\caption{{\bf BioFET-SIM parameters.} The analytical expression for $\lambda_{TF}$ is given in Eq. \ref{eq:lambda_tf}.}
\label{tab:parameters}
\end{table} 

\end{onecolumn}

\balance
\clearpage
\newpage

%\fancyhead{}
%\fancyhead[C]{\footnotesize{ESI - Hediger \emph{et al.} - BioFET-SIM Online}}
%\fancyfoot{}
%%left odd right even
%\fancyfoot[LO,R]{\footnotesize{\sffamily{\emph{PLoS One}, 2012}}}
%\fancyfoot[RO]{\footnotesize{\sffamily{ESI-1--\pageref{LastPage} ~\textbar~\thepage}}}
%\fancyfoot[L]{\footnotesize{\sffamily{\thepage~\textbar ESI-1--\pageref{LastPage}}}}

\begin{onecolumn}
\section*{Electronic Supplementary Information} 
\begin{center} 
%\noi\Large{BioFET-SIM Online} 
%
%\noi\Large{MRH, JHJ, LDV} 

\large{BioFET-SIM Web Interface: Implementation and Two Applications}

Martin R. Hediger, Jan H. Jensen, Luca De Vico

%\author{MRH, JHJ, LDV}
%\author{Martin R. Hediger$^1$\and 
%        Jan H. Jensen$^{1}$\and
%        Luca De Vico$^1$\footnote{luca@chem.ku.dk}
%\small{{}$^1$Department of Chemistry, Copenhagen University, Universitetsparken 5, DK-2100, Denmark}
\end{center} 

%\maketitle

\setcounter{section}{1}
\setcounter{page}{1}
\setcounter{equation}{0}
\setcounter{figure}{0}
\setcounter{table}{0}
\renewcommand{\thesection}{S\arabic{section}}
\renewcommand{\thepage}{ESI-\arabic{page}}
\renewcommand{\theequation}{S\arabic{equation}}
\renewcommand{\thefigure}{S\arabic{figure}}
\renewcommand{\thetable}{S\arabic{table}}

\begin{normalsize} 
\setcounter{section}{1}

\singlespacing

\subsection{Derivation of Relative Nanowire Sensitivity Factor}\label{sec:sup_derv_sens}

According to our model, the sensitivity of a generic $p$-type nanowire, in the single charge approximation, can be expressed as

\begin{equation}
\label{DG/GO}
\frac{\Delta G}{G_0} = - \, \frac{2}{R \, e \, p_0}\Gamma \left( \Gamma_l \, \sigma_b + \sigma_s \right)
\end{equation}

\noindent and the base conductance $G_0$ can be expressed as

\begin{equation}
\label{G0}
G_0 = \frac{\pi \, R^2 \, e \, p_0 \, \mu}{L}
\end{equation}

In these equations $e$ is the elementary charge, $R$ is the radius of the nanowire and $p_0$ the hole density, $\mu$ is the charge carrier mobility.
$\Gamma$ and $\Gamma_l$ are dimensionless functions quantifying the actual sensitivity of the nanowire and they depend, among other parameters, on the distance $l$ of the sensed charge from the nanowire surface and the buffer Debye length $\lambda_D$\cite{de2011quantifying}.
When using a multiple charge model, we interpret $l$ as the average distance of the sensed charges from the nanowire surface.

If the physical and geometrical properties of the nanowire are fixed,
$G_0$ is constant. If we only consider the sensing of charges immersed
in the buffer (i.e. let $\sigma_s = 0$), Eq. \ref{DG/GO} can be
simplified and it is possible to express the change in conductivity as

\begin{equation}
\label{DG}
\Delta G = K \Gamma \, \Gamma_l \, \sigma_b 
\end{equation}

\noindent where $K$ collects all constant values. Using the expression for $\Gamma_l$

\begin{equation}
\label{gamma_l}
\Gamma_l = 2 \frac{R}{R+l} \left(1+\sqrt{\frac{R}{R+l}}\exp(l/\lambda_D)  \right) ^{-1}
\end{equation}

we can now define a value for the buffer Debye length at maximum dilution, $\lambda_{D}^{max}$, and express the
maximum change in conductivity at this value as

\begin{equation}
\label{DGmax}
\Delta G^{max} = K \Gamma^{max} \, \Gamma_l^{max} \, \sigma_b.
\end{equation}

When considering a highly diluted buffer, i.e. $\lambda_{D}^{max} \gg l$, we can express
$\Gamma_l^{max}$ as

\begin{equation}
\label{gamma_lmax}
\Gamma_l^{max} \simeq 2 \frac{R}{R+l} \left(1+\sqrt{\frac{R}{R+l}} \right) ^{-1}.
\end{equation}

The ratio between the change in conductivity at a given Debye length and the maximum possible value becomes

\begin{equation}
\label{DG/DGmax}
\frac{\Delta G}{\Delta G^{max}} =  \frac{K \Gamma \, \Gamma_l \, \sigma_b}{K \Gamma^{max} \, \Gamma_l^{max} \, \sigma_b} = \frac{\Gamma \, \Gamma_l}{\Gamma^{max} \, \Gamma_l^{max}}
\end{equation}

\noindent and after reordering it is possible to obtain

\begin{equation}
\label{tofitl}
\frac{\Delta G \, \Gamma^{max}}{\Delta G^{max} \, \Gamma} = \frac{\Gamma_l}{\Gamma_l^{max}}.
\end{equation}

After inserting the explicit expressions, we obtain

\begin{equation}
\label{tofitr}
\frac{\Gamma_l}{\Gamma_l^{max}} = \frac{2 \frac{R}{R+l} \left(1+\sqrt{\frac{R}{R+l}}\exp(l/\lambda_D)  \right) ^{-1}}{2 \frac{R}{R+l} \left(1+\sqrt{\frac{R}{R+l}} \right) ^{-1}}
\end{equation}

where we define $\Gamma_l/\Gamma_l^{max}$ as the \textit{relative sensitivity factor}.

Using Eq. 2 from previously published work\cite{de2011quantifying}, it is possible to compute the
values of $\Gamma$ for different Debye lengths, including $\Gamma^{max}$
for $\lambda_{D}^{max} = 1000$nm. Using BioFET-SIM it is possible
to obtain the value of $\Delta G$ (and $\Delta G^{max}$) simply by
multiplying $\frac{\Delta G}{G_0}$ with $G_0$. It is then possible to plot the l.h.s. of Eq. \ref{tofitl} for different values of
$\lambda_D$. This plot can be fitted to the r.h.s. of
Eq. \ref{tofitr} where $l$ is the fitting paramenter. From a series of measures at different Debye lengths, it is then possible to obtain the
average distance of the sensed charge from the nanowire surface.

\subsection{Expression for $\Gamma$}\label{subsec:gamma}

For simplification of notation, in the expression for $\Gamma$, we use the thickness $t := R_{NW} + l_{ox}$.
Then, $\Gamma$ is given by

\begin{equation}
\label{eq:gamma}
\Gamma = \frac{\epsilon_1 \cdot K_0\left(\frac{t}{\lambda_D}\right)\frac{\lambda_D}{\lambda_{TF}} \cdot I_1\left(\frac{R_{NW}}{\lambda_{TF}}\right)}{\left[ K_0\left(\frac{t}{\lambda_D}\right) \cdot \frac{\lambda_D}{t} + ln\left(\frac{t}{R_{NW}}\right) \cdot K_1\left(\frac{t}{\lambda_D}\right) \frac{\epsilon_3}{\epsilon_2} \right] \epsilon_1 \frac{R_{NW}}{\lambda_{TF}} \cdot I_1\left(\frac{R_{NW}}{\lambda_{TF}}\right) + \epsilon_3 \cdot K_1\left(\frac{t}{\lambda_D}\right) \cdot I_0\left(\frac{R_{NW}}{\lambda_{TF}}\right)}
\end{equation}

In Eq. \ref{eq:gamma}, $I_0$, $I_1$, $K_0$ and $K_1$ are the modified Bessel functions of first and second kind, respectively\cite{de2011quantifying}.

\subsection{Antibody and Antigen Preparation}\label{sec:sup_ab_prep}

A suitable complex structure of a generic antibody and an antigen used by Vacic et al.\cite{ja205684a} was prepared.
Only few full antibody structures have been resolved. We used the structure of an intact IgG2a monoclonal antibody, ascension code 1IGT\cite{harris1997refined}.
For the antigen we used the structure of the SEA domain of human mucin 1, with ascension code MUC1\cite{macao2005autoproteolysis}.

The antigen structure was rigidly docked to the antigen-binding site of
the antibody using AutoDock\cite{goodsell1996automated} and visually
checked with the program PyMOL\cite{PyMOLu} for a reasonable docking. The
scope of this docking was only to obtain a feasible complex structure.

Since we were interested in the sensing of only the antigen, the antibody structure had to be made as neutral as possible.
In order to make the antibody as neutral as possible, all positions in the antibody sequence were mutated to glycine using PyMOL. When BioFET-SIM computes the charges of a protein, a positive charge is assigned to the N-terminus and a negative charge to the C-terminus, depending on the corresponding pK$_a$ values calculated by PROPKA\cite{ct100578z}. In order to counter balance the charges of the termini, in each of the four chains of the antibody we mutated the residue at the N-termini to aspartate and the residue at the C-termini to arginine.

\subsection{Antigen Sensing}

The BioFET-SIM parameters were set as in Tab. \ref{tab:parameters}. With these parameters we obtain $G_0 = 279.0$ nS.

The different orientations of the neutral antibody/antigen model complex which were tested are shown in Fig. \ref{fig:orientations}.
A biofunctionalization layer of 0.5 and 1.0nm was added for C- and N-terminus binding to the nanowire surface, respectively.
We considered a pH of 7.4 and kept the number of proteins fixed to 4000 protein units.
The BioFET-SIM results for all orientations A-G at the values of Debye length employed by Vacic et al. (3.07, 9.7 and 30.7nm) and $\lambda_{D}^{max} = 1000$nm are reported in Tab. \ref{tablerawdata}, together with the corresponding values for
$\Gamma$.
In Tab. \ref{tabletofit}, the derived values for the l.h.s. of Eq. \ref{tofitl} are reported.
The data from Tab. \ref{tabletofit} is plotted in Fig. \ref{tot_orient}, together with the fitting parameter.

\clearpage

%\begin{figuresupp}[htb!]
\begin{figure}[htb!]
\caption{Different orientations of the neutral antibody/antigen
complex. In orientations A, C and G, the complex is bound by the C-termini, in orientations B, D, E and F, the complex is bound by the N-terminus.}
\label{fig:orientations}
\begin{center}
\begin{tabular}{c c c c}
A) & \includegraphics[width=5cm]{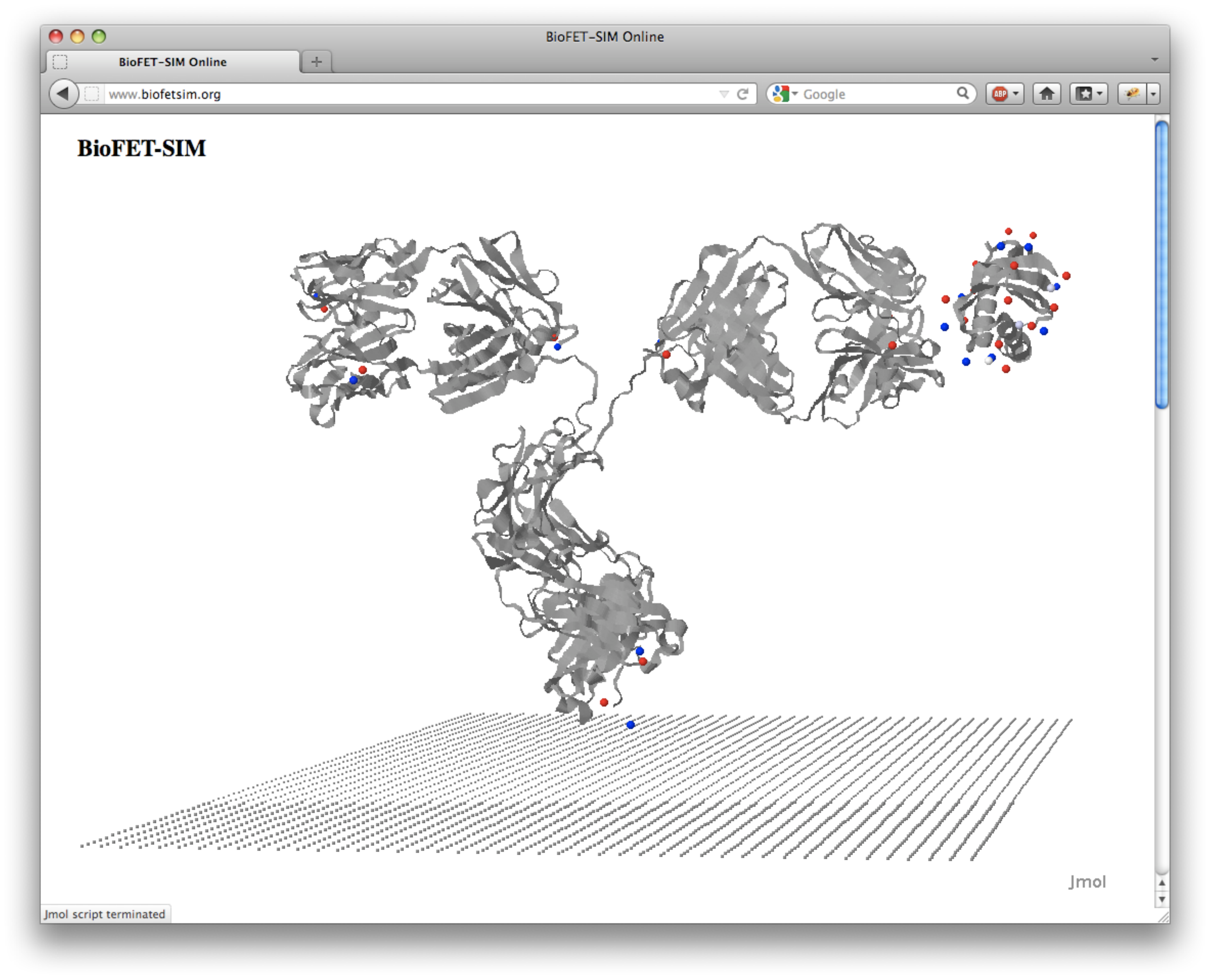} & B) & \includegraphics[width=5cm]{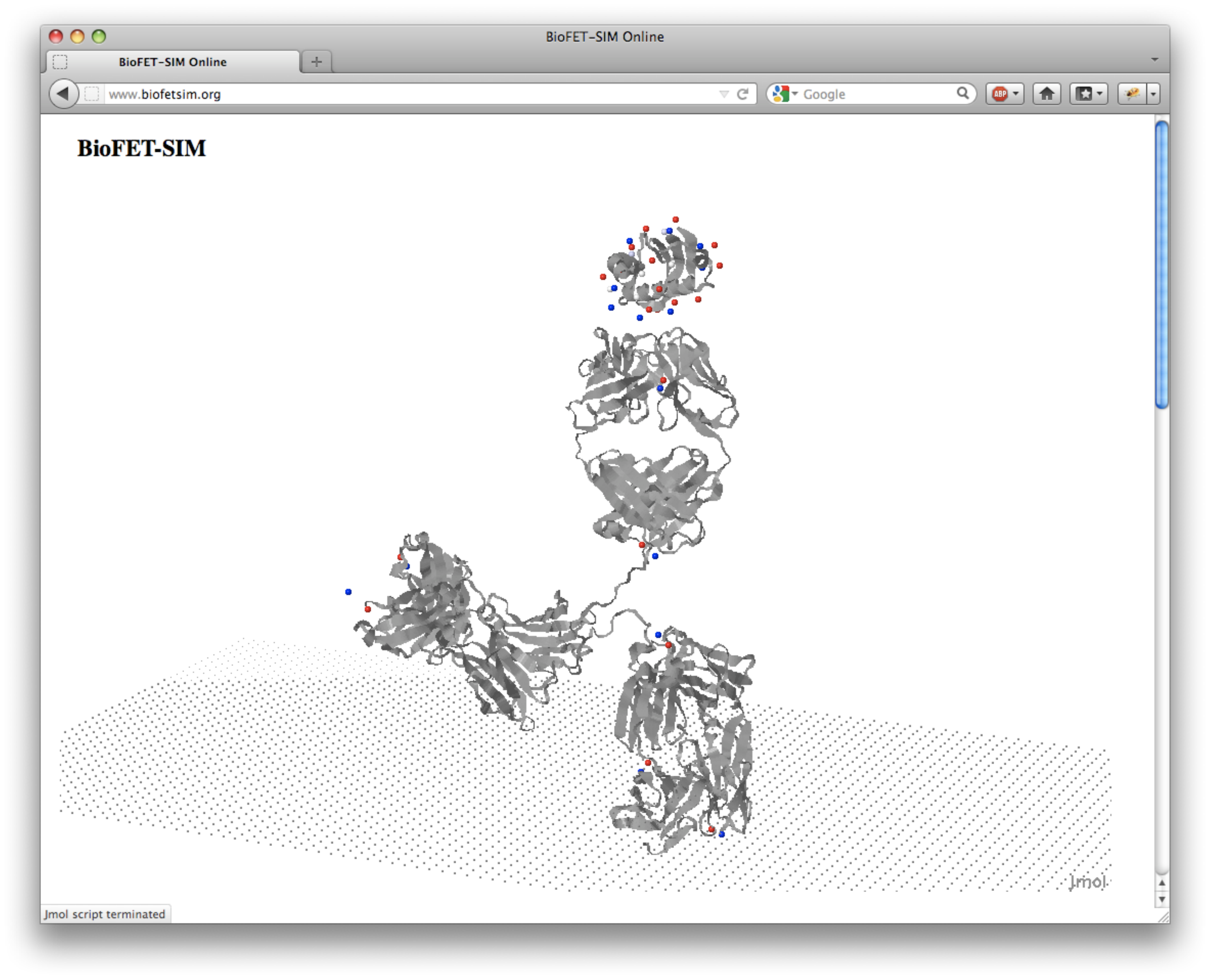} \\
C) & \includegraphics[width=5cm]{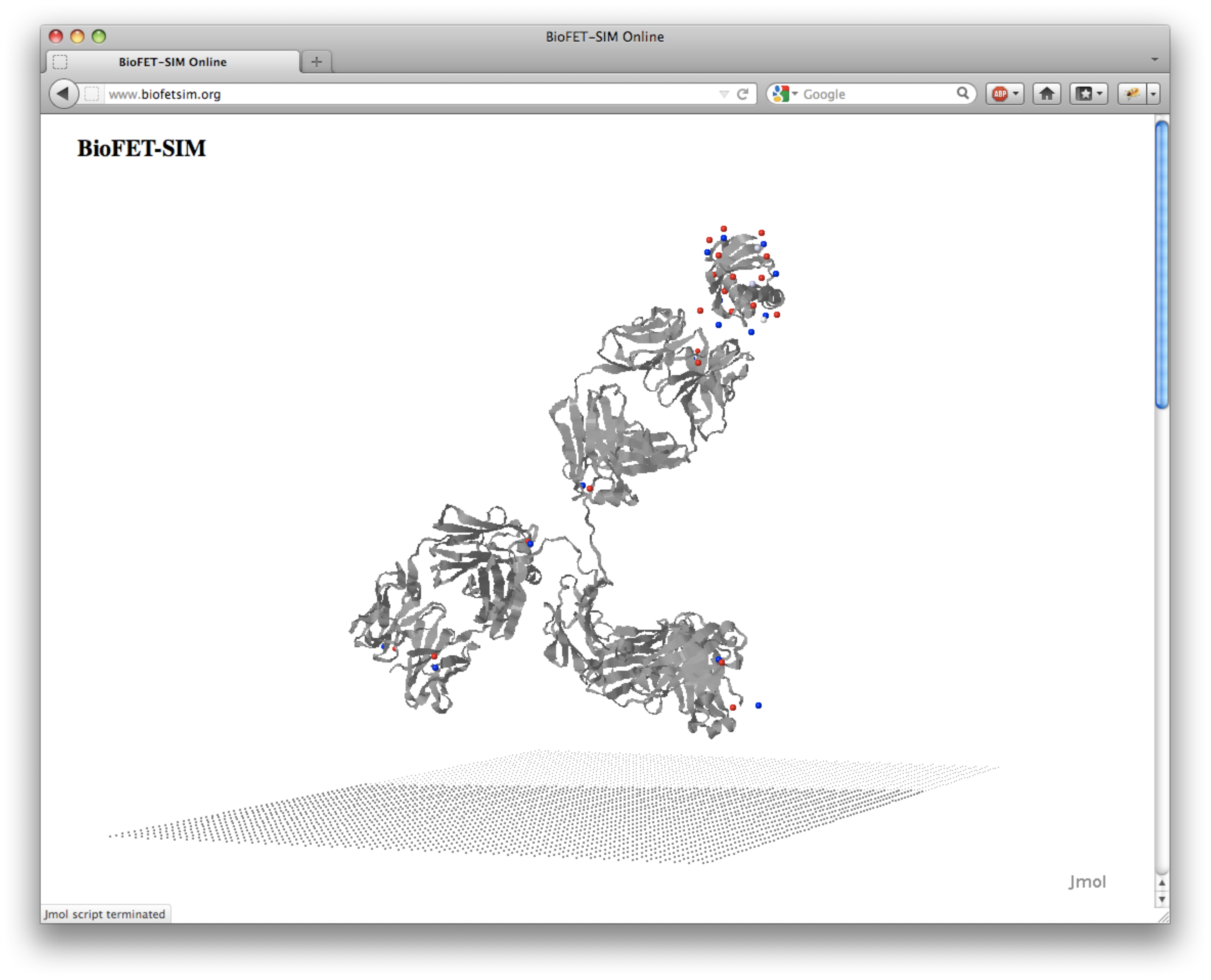} & D) & \includegraphics[width=5cm]{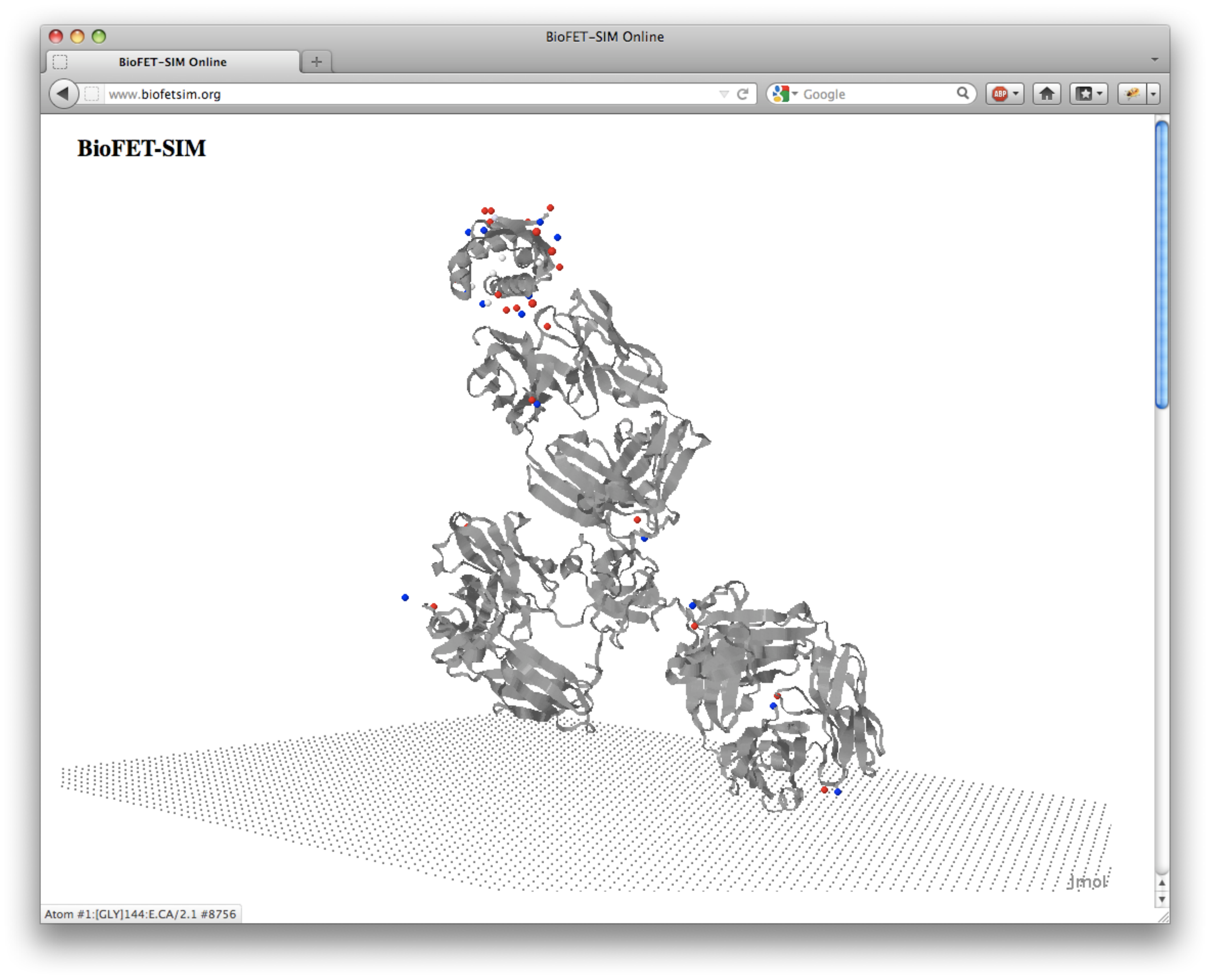} \\
E) & \includegraphics[width=5cm]{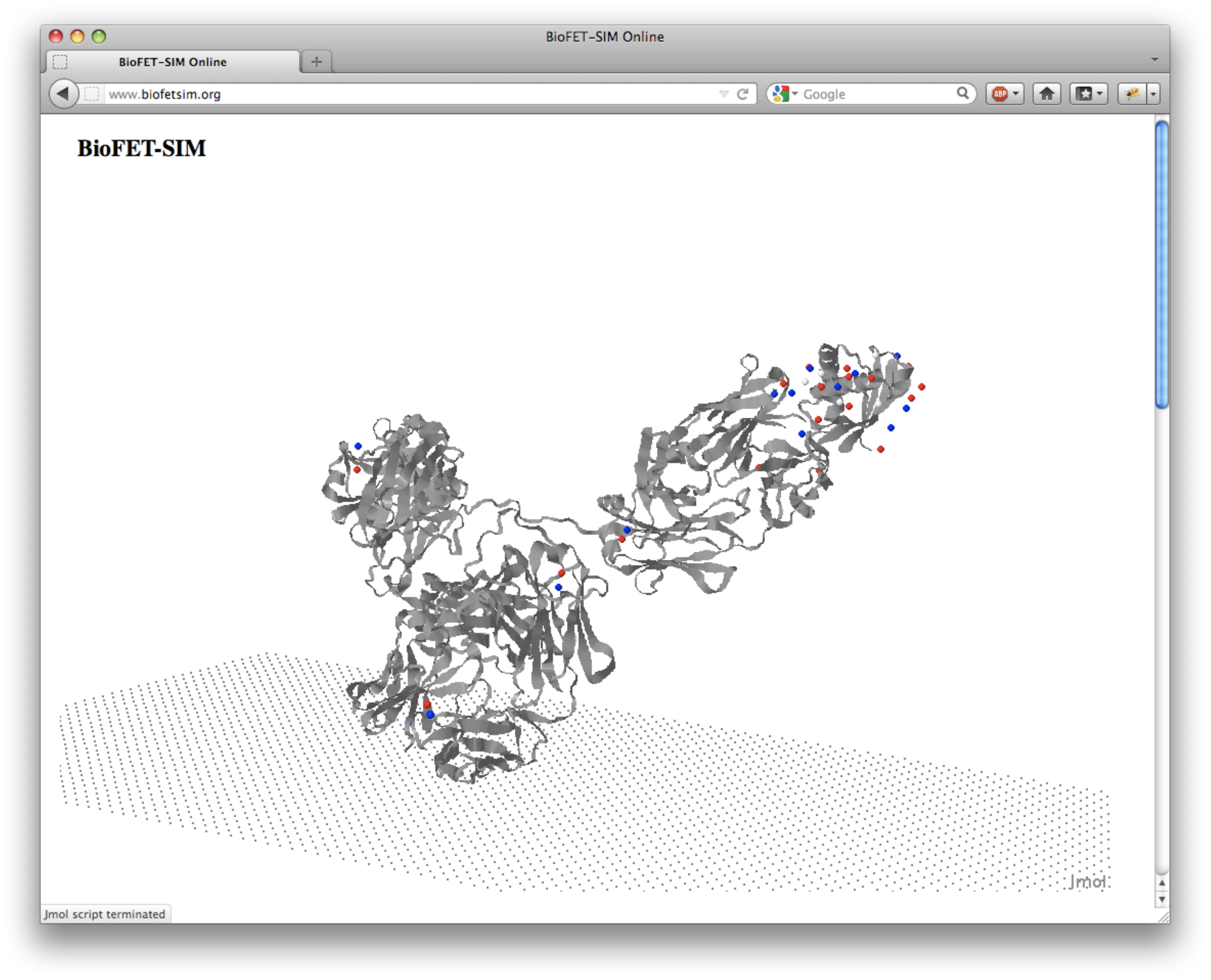} & F) & \includegraphics[width=5cm]{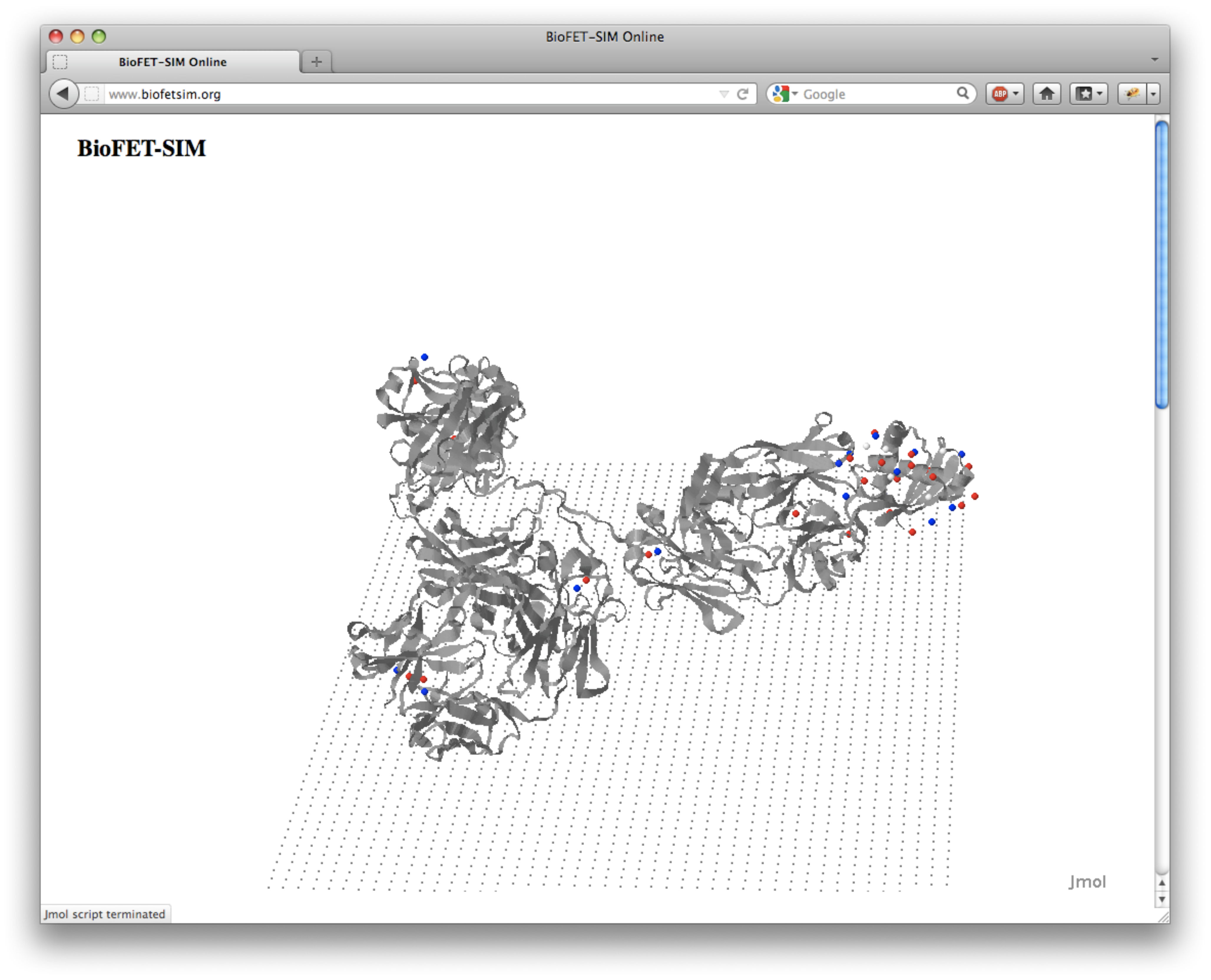} \\
G) & \includegraphics[width=5cm]{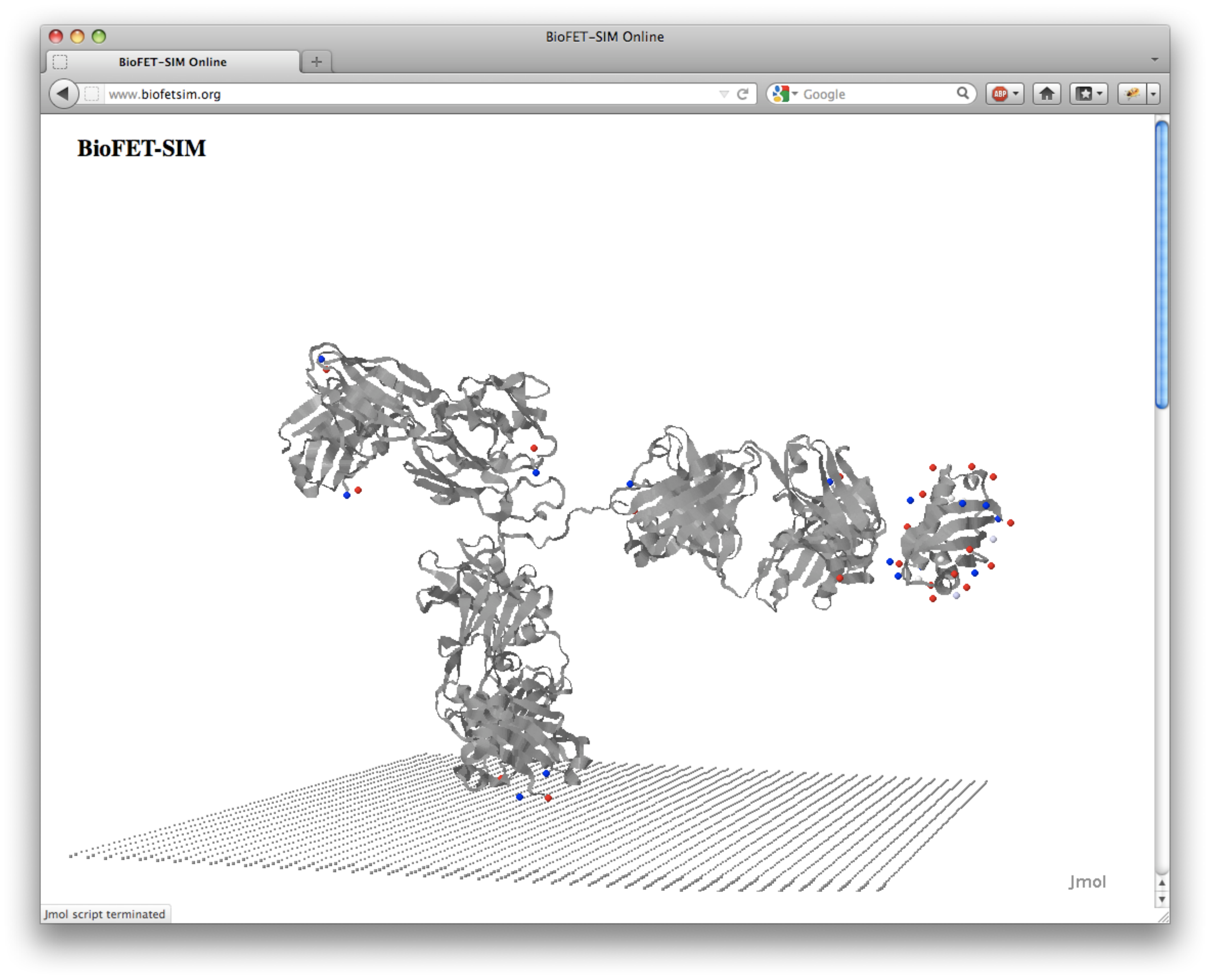} & & \\
\end{tabular}
\end{center}
\end{figure}
%\end{figuresupp}

\clearpage
%\begin{tablesupp}[htb!]
\begin{table}[htb!]
\caption{Sensitivity}
\label{tablerawdata}
\begin{center}
\scriptsize
\begin{tabular}{c | c c c c c c c | c}
Orientations&A&B&C&D&E&F&G&\\
Debye length [nm] &$\frac{\Delta G}{G_0}$&$\frac{\Delta G}{G_0}$&$\frac{\Delta G}{G_0}$&$\frac{\Delta G}{G_0}$&$\frac{\Delta G}{G_0}$&$\frac{\Delta G}{G_0}$&$\frac{\Delta G}{G_0}$&$\Gamma$\\
\hline
3.07&0.005466&0.006495&0.019521&0.040992&0.120386&0.254474&0.056443&0.0514\\
9.7&0.549386&0.260311&0.403622&0.666037&1.11689&1.574997&0.877657&0.1257\\
30.7&1.866226&1.161079&1.508332&2.083992&2.981374&3.83572&2.529128&0.2403\\
1000.0&5.24021&3.506654&4.356305&5.737461&7.834166&9.786054&6.795953&0.5651\\
\end{tabular}
\end{center}
\end{table}
%\end{tablesupp}

%\begin{tablesupp}[htb!]
\begin{table}[htb!]
\caption{Relative sensitivity factor, values for orientations A-G.}
\label{tabletofit}
\begin{center}
\scriptsize
\begin{tabular}{c | c c c c c c c}
Debye length [nm] &A&B&C&D&E&F&G\\\\
\hline
3.07&0.01146788&0.020363313&0.049265849&0.078549161&0.168945023&0.285889363&0.09131075\\
9.7&0.471323318&0.333725685&0.416530283&0.521877579&0.640925665&0.723539345&0.580583226\\
30.7&0.837504368&0.778646598&0.814235818&0.85417693&0.894942693&0.921745075&0.875169108\\
\end{tabular}
\end{center}
%\end{tablesupp}
\end{table}

%\begin{figuresupp}[htb!]
\begin{figure}[htb!]
\caption{Fit of relative sensitivty factor against data from Tab. \ref{tabletofit}.}
\label{tot_orient}
\includegraphics[width=9cm]{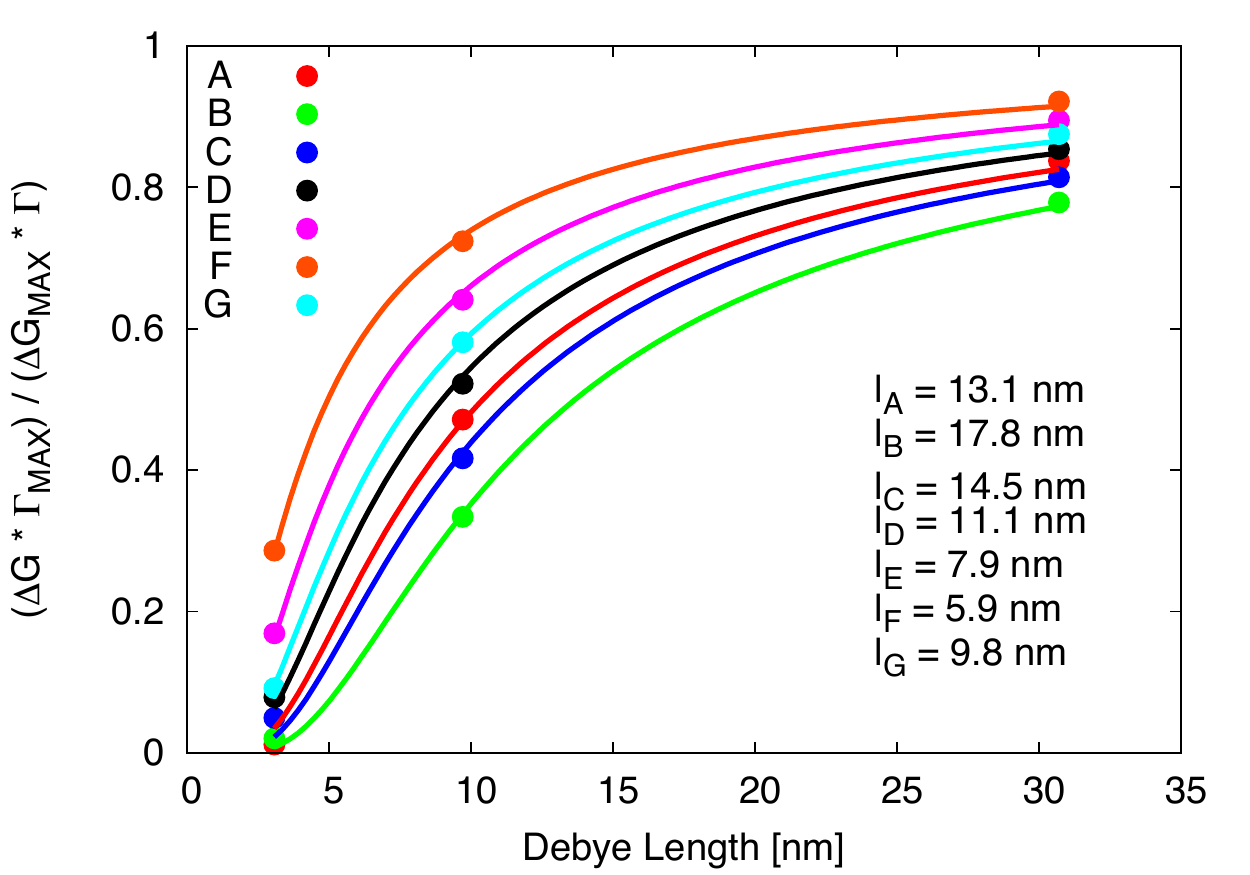}
\end{figure}
%\end{figuresupp}

\onecolumn 
\newpage
\subsection{Command Line Version of BioFET-SIM Usage}\label{sec:sup_biof_comd}

PDB Format Requirements:\\
The uploaded PDB file is required to contain the \verb+MODEL+, \verb+TER+ and \verb+END+ tags. In addition, individual chains of the structure are required to contain the chain label (i.e. \verb+A+, \verb+B+, ...).
\begin{verbatim}
MODEL        1
ATOM      1  N   ASN A   2       0.209  -1.748  -0.613  1.00  0.00
...
TER
ATOM      1  N   HIS B   2      12.057   2.821   9.469  1.00  0.00 
...
END
\end{verbatim}

A number of instruction videos are available, the links to these are found on the interface website, www.biofetsim.org.
The link to the command line version repository is found at the same URL.

\newpage

System Requirements:\\
The command line version of the BioFET-SIM program requires the following applications and libraries to be installed on the local host.
\begin{itemize}
    \item Python 2.5 or higher (not Python 3)
    \item Numpy and Scipy libraries installed
\end{itemize}

Basic usage is demonstrated by issuing
\begin{verbatim}
[user] $ python bio_run.py 
BioFET-SIM usage:
$ python bio_run.py --calc <input.bfs>
or
$ python bio_run.py --set <param> <val> <input.bfs>
\end{verbatim} 

Starting a BioFET-SIM calculation using the input file "kk8add.bfs":
\begin{verbatim}
[user] $ python bio_run.py --calc kk8add.bfs
# BioFET-SIM Calculation
# Date of calculation: 2012-03-21 17:17:51
# Calculation target: kk8add
# pH: 7.4
# Comment: BFS Input generated by interface.
                                    
Adjustable Parameters:
L_d                        2.0
L_tf                      2.04
eps_1                     12.0
eps_2                      3.9
eps_3                     78.0
lay_bf                     1.0
lay_ox                     2.0
mu                        0.01
n_0                   1.11e+24
num_prot                  4000
num_qi                       6
nw_len                  2000.0
nw_rad                    10.0
nw_type                      P
target                  kk8add

Base Conductance [nS]: 279.352916413
Sensitivity:           0.115675065297
\end{verbatim}

In the output, the labels have the following meaning (with the corresponding symbol given in Tab. \ref{tab:parameters}, $q_i$ given in Eq. \ref{eq:charge_pH}): \verb+L_d+: Debye screening length $\lambda_D$; \verb+L_tf+: Thomas-Fermi screening length $\lambda_{TF}$; \verb+eps_1+: Nanowire permittivity $\epsilon_1$; \verb+eps_2+: Oxide layer permittivity $\epsilon_2$; \verb+eps_3+: Solvent permittivity $\epsilon_3$; \verb+lay_bf+: Biolayer thickness $l_b$; \verb+lay_ox+: Oxide layer thickness $l_{ox}$; \verb+mu+: Charge carrier mobility $\mu$; \verb+n_0+: Charge carrier density $\kappa_0$; \verb+num_prot+: Number of biomolecules $N$; \verb+num_qi+: Number of charged sites within each biomolecule $q_i$; \verb+nw_len+: Nanowire length $L_{NW}$; \verb+nw_rad+: Nanowire radius $R_{NW}$; \verb+nw_type+: Nanowire doping type $K$; \verb+target+: PDB identifier of the studied biomolecule.

\newpage
A parameter can be adjusted by the "--set" option, followed by the label for the corresponding parameter and the new value:
\begin{verbatim}
[user] $ python bio_run.py --set L_d 3.0 kk8add.bfs
Parameter adjusted.
\end{verbatim} 

%\end{document}

%\bibliographystyle{achemso}
\bibliographystyle{001-arxiv_2009}
\bibliography{citations}

\end{normalsize}
\end{onecolumn}

\end{document}